\documentclass[12pt]{article}

\usepackage{amsmath,amsthm,amssymb,euscript,epsf,epsfig}
\usepackage{array}
\usepackage{fancybox}
\usepackage{comment} 
\usepackage{cite}
\usepackage{rotating}

\usepackage{tikz,pbox}
\usetikzlibrary{shapes,arrows}
\usetikzlibrary{calc,decorations.markings}

\usepackage{hyperref}

\makeatletter
\@addtoreset{equation}{section}
\makeatother

\def\one{{\hbox{ 1\kern-.8mm l}}}
\def\zero{{\hbox{ 0\kern-1.5mm 0}}}

\def\mC{ \mathbb{C}}

  \def\cC{{\cal C}}
  
 \def\cH{{\cal H}}

 \def\cZ{{\cal Z}}

\newcommand{\be}{\begin{equation}}
\newcommand{\ee}{\end{equation}}
\newcommand{\beq}{\begin{equation}}
\newcommand{\eeq}{\end{equation}}
\newcommand{\bea}{\begin{eqnarray}\displaystyle}
\newcommand{\eea}{\end{eqnarray}}

\newcommand{\tr}{{\rm tr}}

\begin{document}

\begin{flushright}
QMUL-PH-21-54
\end{flushright}

\medskip

\begin{center}

{ \Large\bf  Combinatoric topological string theories \\ and group theory algorithms }

\bigskip

Sanjaye Ramgoolam$^{a,b,\dagger }$ and Eric Sharpe$^{c , \dagger \dagger } $

\bigskip

{\small

$^{a}$
{\em Centre for Theoretical Physics, Department of  Physics,}\\
{\em Queen Mary University of London, London E1 4NS, United Kingdom }\\
\medskip
$^{b}${\em  School of Physics and Mandelstam Institute for Theoretical Physics,}\\
{\em University of Witwatersrand, Wits, 2050, South Africa} \\
\medskip
${}^c ${ \em Department of Physics MC 0435, 850 West Campus Drive }\\
{ \em Virginia Tech, Blacksburg, VA 24061}\\
}

\begin{abstract}

A number of finite algorithms for constructing representation theoretic 
data from group multiplications in a finite group $G$   have recently been shown to be related to   
amplitudes for combinatoric topological strings ($G$-CTST) based on Dijkgraaf-Witten theory of flat $G$-bundles on surfaces.  We extend this result to projective representations  of $G$ using twisted Dijkgraaf-Witten theory. 
New algorithms for characters are described, based on handle creation operators and minimal multiplicative generating subspaces for the centers of group algebras and twisted group algebras. Such minimal generating subspaces are of interest in connection with  information theoretic aspects of the AdS/CFT correspondence. For the untwisted case, we describe the integrality properties of certain character sums  and character power sums which follow from these constructive $G$-CTST algorithms. These integer sums appear as residues of singularities in $G$-CTST generating  functions.   $S$-duality of the combinatoric topological strings motivates the definition of an inverse handle creation operator in the centers of group algebras and twisted group algebras.

\end{abstract}

\end{center}

\vfill 

{\small{ 
\noindent 
 { E-mails: ${}^{\dagger }$s.ramgoolam@qmul.ac.uk , 
${}^{ \dagger \dagger}$ersharpe@vt.edu  } 
} }

\newpage 

\tableofcontents

\section{Introduction  }

Two-dimensional Dijkgraaf-Witten theories are simple examples of topological
field theories associated to finite groups
\cite{Dijkgraaf:1989pz,Witten:1991we,FQ,FHK}.  At a basic level,
these theories describe orbifolds of points, $[{\rm point}/G]$, possibly with
discrete torsion (described in this context as a twisting).
In the case when the group
is a symmetric group $S_n$, these theories admit defects, which have applications in describing counting and correlators in  $U(N)$ gauge theories \cite{QuivCalc} 
of interest in AdS/CFT \cite{malda,GKP,WittenHol}. 
Recent work on wormhole physics and baby universes  
\cite{MarMax,Gardiner:2020vjp,deMelloKoch:2021lqp,Couch:2021wsm,Moore,Heckman:2021vzx,Schlenker:2022dyo}, 
in the context of topology change in quantum gravity, 
considers sums over Riemann surfaces weighted by a string coupling $g_{st}$, 
where each surface supports a Dijkgraaf-Witten theory.
We will refer to these theories, summing over worldsheets,
as combinatoric topological string theories or $G$-CTST. Motivations and insights on the mathematical properties of these strings thus arise both from AdS/CFT and from models of topology change in quantum gravity.
 
Another place Dijkgraaf-Witten theories arise is in couplings
to physical theories.  For example, consider an orbifold $[X/\Gamma]$,
where a subgroup $K \subset \Gamma$ acts trivially on $X$, as studied
in e.g.~\cite{Pantev:2005rh,Pantev:2005wj,Pantev:2005zs,Hellerman:2006zs,Robbins:2020msp,Robbins:2021lry,Robbins:2021ibx,Robbins:2021xce,Sharpe:2021srf}.
This can be interpreted as a coupling of the orbifold $[X/G]$ (for
$G = \Gamma/K$) to Dijkgraaf-Witten theory for the group $K$,
as will be discussed in greater detail in \cite{es-toappear}.
The orbifold $[X/\Gamma]$ is in any event equivalent to
a disjoint union of orbifolds, a result known as decomposition 
\cite{Hellerman:2006zs}, which when viewed as a coupling of a topological field
theory, reflects the fact that as a topological field theory,
Dijkgraaf-Witten theory itself
is a disjoint union of invertible field theories
\cite{Durhuus:1993cq,Moore:2006dw,Komargodski:2020mxz,Huang:2021zvu}.
Applied to $G$-CTST, decomposition implies that the `string field theory'
of Dijkgraaf-Witten theory (in the same sense as \cite{Witten:1992fb})
is a theory on a disjoint union of points, which 
could be interpreted as a noninteracting statistical mechanical theory.

In the recent paper \cite{deMelloKoch:2021lqp} it was observed that well-known formulae for  
amplitudes in $G$-CTST can be used to give a finite algorithm which starts from 
 group multiplications in $G$ and arrives at  the integer ratios ${ |G|/{ ( \dim R ) } }$ (relating the order of a finite group $G$ and the dimension of an irreducible representation $R$). 
The integrality of these ratios is 
an interesting old result at the intersection of finite group theory and number theory (see for example \cite{Schur,Simon}) 
and plays an important role in the algorithm. 
The form of the group multiplications in the input is understood geometrically in terms of the fundamental groups of two dimensional surfaces, which are interpreted in $G$-CTST as string worldsheets. The algorithm proceeds by finding the zeroes of a polynomial equation which has integer coefficients (which are $G$-CTST amplitudes) and has roots which are also known to be integers (i.e. the ${ |G|/{ ( \dim R ) } }$ ). The construction of representation theoretic quantities using combinatoric methods is an interesting general theme in representation theory \cite{BarcRam}, with implications for computational complexity theory \cite{PakPanova,IkMulWal}. $G$-CTST provides an interesting topological perspective on this theme.  A quantum mechanics of bipartite ribbon graphs which constructs Kronecker coefficients as eigenvalue degeneracies of Hamiltonians \cite{JBSRKron} is another angle on the theme of exploiting stringy geometric/algebraic structures to address questions in combinatorial representation theory.

It is natural to consider twisted $G$-CTST (involving Dijkgraaf-Witten
theories of orbifolds with discrete torsion)
and its relation to the combinatorics of projective representations of $G$. 
In this paper we will show how the amplitudes in the  
vacuum sector of $G$-CTST  
can be used to obtain the integer ratios ${ |G|/(\dim R ) }$,  
where $\dim R$  is the dimension of a projective representation $R$.  The algorithm takes as input group multiplications weighted by cocycle factors defining the twist, and proceeds by solving a polynomial equation as in 
\cite{deMelloKoch:2021lqp}. 
(The fact that these ratios are always integers in the projective case is 
proven in  \cite{Schur},
\cite[theorem 3.5]{Chen}.)

Standard algorithms for the construction of characters were also shown in \cite{deMelloKoch:2021lqp}  to be related to amplitudes in $G$-CTST, for two dimensional surfaces with boundary circles. In this paper we show that the geometrical picture based on $G$-CTST, along with the study of generating subspaces of centers of group algebras \cite{KempRam}, can be used to give new algorithms for characters. 
The handle creation operator of $G$-CTST plays a role in one class of such algorithms. 
An interesting corollary of this discussion is that string amplitudes with one boundary in  $G$-CTST determine a distinguished subspace of the center of
the twisted group algebra, $ \cZ ( \mC_{\omega} ( G ) )$,  of dimension equal to the number of distinct integers $\dim R$. 
This discussion will be presented for both the untwisted and the twisted case. 

The study of generating subspaces of centers of  symmetric group algebras in \cite{KempRam} was motivated by the consideration of  a toy model for black hole information loss arising from the AdS/CFT correspondence \cite{BCLS}. A family of supergravity solutions \cite{LLM} with 
$AdS_5 \times S^5$ asymptotics are dual 
to half-BPS states in the dual  CFT labelled by Young diagrams \cite{CJR}.  As explained in \cite{BCLS} the  asymptotic gravitational charges of the SUGRA solutions correspond to Casimirs of the $U(N)$ gauge symmetry in the CFT. The information loss model considers the information content in a finite number of Casimirs. For quantum states having  energy $n$ in the natural units, the Casimirs are related by Schur-Weyl duality to central elements in the group algebra of $\mC ( S_n ) $. The information content in low order  Casimirs translates into a question about how effectively low order cycle operators in the center of $\mC ( S_n)$ distinguish Young diagrams. This is in turn related to the dimensions of subspaces of the center generated by a finite set of central elements. 
In this paper we will be considering the generating subspaces for general finite groups $G$ in connection with Dijkgraaf-Witten topological field theories. The embedding of this discussion into gauge-string dualities is an interesting problem  for the  future.

The paper is organised as follows. Section \ref{sec:FTVac} explains the use of amplitudes in the vacuum sector of $G$-CTST to give finite algorithms starting from group multiplications in $G$  weighted by appropriate cocycle factors and deriving the integer ratios ${ |G|/\dim  R } $ for projective representations $R$ of finite groups $G$. The handle creation operator \eqref{eq:handle-op-id} for twisted group algebras plays an important role in this discussion. By considering one-point functions of twist field operators on higher genus surfaces, expressible combinatorially using the handle creation operator,  we give a combinatoric construction for the number of distinct dimensions $ \dim R $ for irreducible representations of $G$, or irreducible projective representations of $G$. 
Section \ref{sec:CharBdy} extends the discussion to amplitudes in $G$-CTST for surfaces having boundaries to obtain algorithms for calculating characters.  The constructions in sections \ref{subsecPismall},\ref{sec:char-gen}, \ref{sec:fin-min-gen} are used to obtain some integrality properties of certain sums of characters and sums of powers of characters in section \ref{sec:propchar}, which in turn have implications for factorisation properties of certain polynomials which are used in character algorithms \cite{Burnside,Dixon,Schneider}.  The integer sums and power sums 
of characters appear as residues for singularities in appropriate $G$-CTST partition functions.  For simplicity this section focuses on the untwisted case. Section \ref{sec:GCTST} collects a few remarks on $G$-CTST: we elaborate on the connection between  determinants appearing in the algorithms of sections [\ref{sec:FTVac},\ref{sec:CharBdy}] and plethystic exponentials of stringy amplitudes at low genus. We also comment on $S$-duality in $G$-CTST, which leads to the definition of an inverse handle creation operator. This is given as an expansion in terms of the projector basis of $\cZ( \mC_{\omega} ( G ) )$, while its expansion in terms of the conjugacy class basis is an interesting question for the future.

\section{Fourier transform and vacuum sector for $G$-CTST}\label{sec:FTVac} 

The previous paper \cite{deMelloKoch:2021lqp} studied computations of characters
of ordinary representations of finite groups, as relevant to e.g.~the
AdS/CFT correspondence.  In this section we generalize those
computations to include discrete torsion, which twists the representations
to projective representations.  In broad brushstrokes, much of the
analysis is formally similar to \cite{deMelloKoch:2021lqp}, so we will combine a review of
the results of \cite{deMelloKoch:2021lqp} while simultaneously describing novel features
present in cases with discrete torsion.

To improve readability, we have banished a number of technical
definitions and computations in cases with discrete torsion to
appendix~\ref{app:2ddw-results}, to which we refer as needed.

\subsection{The twisted group algebra of a finite group $\mC_{\omega} ( G )$}

Let $G$ be a finite group and
$[\omega] \in H^2(G,U(1))$.  In this section we will review properties
of the twisted group algebra $\mC_{\omega}(G)$ and its center
$\cH = \cZ ( \mC_{\omega} ( G ) )$, which will play an important role
in our computations.  Physically, the center $\cH$ is the state space
of a two-dimensional (twisted) Dijkgraaf-Witten theory, which
we will call $G$-CTST for short. Setting $\omega =1$ in the formulae that follow recovers formulae for centers of ordinary group algebras $\cZ ( \mC ( G ) )$.

The twisted group algebra
$\mC_{\omega} ( G )$ is a vector space, with basis elements we label
$\tau_g$ corresponding to
elements $g$ of the group $G$,
equipped with the product 
\begin{equation}
\tau_g \tau_h \: = \: \omega(g,h) \tau_{gh},
\end{equation}
which is generically non-commutative.  $\omega$ is a $2$-cocycle representing the cohomology class $[\omega]$. Generic elements take the form 
\bea 
\sum_{ g\in G  } a_g \tau_g ,
\eea
where $a_g \in \mC$. 
$\mC_{\omega} ( G )$ has an inner product where the group elements are orthonormal: 
\bea
\langle g_1 | g_2 \rangle \: = \: \delta ( g_1 g_2^{-1} ) . 
\eea
For general elements 
\bea\label{innprod}  
\left\langle \sum_{ g_1 } a_{ g_1 } \tau_{g_1} \Bigg|
 \sum_{ g_2} b_{g_2} \tau_{g_2} \right\rangle 
\: = \:
 \sum_{ g_1 , g_2 } a_{ g_1 }^*  b_{ g_2} \delta ( g_1 g_2^{-1} ) .
\eea

Now, we are interested in the center of $\mC_{\omega} ( G )$,
denoted $\cH$ earlier, which is the subspace of $\mC_{\omega} ( G ) $
which commutes with $\tau_g$ for  any  $g \in G$.
It inherits an inner product from $\mC_{\omega} ( G )$ by restriction 
of~\eqref{innprod}. 
One basis for the center is given by twist fields, which are associated with 
$\omega$-regular conjugacy classes.
An element $g \in G$
is said to be $\omega$-regular if for all $h$ commuting with $g$,
\begin{equation}
\omega(g,h) \: = \: \omega(h,g),
\end{equation}
and an $\omega$-regular conjugacy class
is defined \cite[section 3.6]{karpilovsky}
to be a conjugacy class in which every element is $\omega$-regular.

Given an $\omega$-regular conjugacy class $[g]$ represented by
$g \in G$, we define a twist field \cite{karpilovsky}
\begin{eqnarray} 
T_{[g]} & = &
\frac{1}{|G|} \sum_{h \in G}
 \tau_h \tau_g \tau_h^{-1},
\\
& = &  \label{TpCp}
\frac{1}{|G|} \sum_{h \in G} 
\frac{ \omega(h,g) \, \omega(hg,h^{-1}) }{ \omega(h,h^{-1}) }
\tau_{hgh^{-1}}. 
\end{eqnarray}
It can be shown (see for example
\cite[section 2.2.1]{Sharpe:2021srf}) that the twist fields
commute with all elements of
the twisted group algebra $\mC_{\omega} ( G )$, meaning
\begin{equation}
T_{[g]} \tau_h \: = \: \tau_h T_{[g]}
\end{equation}
for all $h \in G$,
and also the $\{ T_{[g]} \}$ form a basis
for the center.

Note that these operators $T_{[g]}$ depend upon the representative
$g$ of the conjugacy class:  as shown in e.g. \cite[section 3]{karpilovsky},
\begin{equation} \label{eq:twist-conj}
T_{[hgh^{-1}]} \: = \: \frac{ \omega(gh^{-1},h) }{ \omega(h,gh^{-1}) }
\, T_{[g]}.
\end{equation}

There is a second basis for the center, given by projectors associated
to irreducible projective representations, which are in (noncanonical)
one-to-one correspondence with $\omega$-regular conjugacy classes.  
(Thus, there are as many projectors as twist fields.)
Let us review some pertinent
results on projective representations before defining those projectors.

Projectors will be constructed using characters of projective representations.
Unlike characters of ordinary representations,
characters of projective representations 
are not class functions, as they are not invariant under conjugation.
If $R$ is a projective representation of
$G$, associated to some cocycle $\omega$, and $\chi^R$ denotes the
character, then \cite[section 7.2, prop. 2.2]{karpilovsky}
\begin{equation} \label{eq:conj-char}
\chi^R(g) \: = \: \frac{ \omega(g,h^{-1}) }{ 
\omega(h^{-1}, h g h^{-1}) }
\chi^R(h g h^{-1}).
\end{equation}

As a consistency check, it may be useful to note that
\begin{eqnarray}
\chi^R\left(T_{[g]}\right)
& = &
\frac{1}{|G|} \sum_{h \in G} \frac{ \omega(h,g) \, \omega(hg,h^{-1}) }{
 \omega(h,h^{-1}) }
\chi^R\left( hgh^{-1} \right),
\\
& = &
\frac{1}{|G|} \sum_{h \in G} \frac{ \omega(h,g) \, \omega(hg,h^{-1}) }{
 \omega(h,h^{-1}) }
\frac{ \omega(h^{-1}, h g h^{-1}) }{ \omega(g,h^{-1}) }
\chi^R(g),
\\
& = &
\frac{1}{|G|} \sum_{h \in G} \chi^R(g) \: = \: \chi^R(g),
\end{eqnarray}
using the fact that
\begin{eqnarray}
\frac{ \omega(h,g) \, \omega(hg,h^{-1}) \, \omega(h^{-1}, hgh^{-1}) }{
\omega(h,h^{-1}) \, \omega(g,h^{-1}) 
}
& = &
 (d\omega)(h^{-1},hg,h^{-1}) \, 
(d\omega)(h^{-1},h,g) \,
\nonumber \\
& & \hspace*{0.5in} \cdot
(d\omega)(h,h^{-1},h),
\nonumber \\
& = & 1.
\end{eqnarray}

In fact, using this identity, one can show
\begin{equation}
\tau_h \tau_g \tau_h^{-1} \: = \: \frac{ 
\omega(h,g) \, \omega(hg,h^{-1}) }{ \omega(h,h^{-1}) } 
\tau_{hgh^{-1}} 
\: = \: \frac{ \omega(g,h^{-1}) }{ \omega(h^{-1}, h g h^{-1}) }
\tau_{h g h^{-1}},
\end{equation}
so we can write~(\ref{eq:conj-char}) as
\begin{equation}
\chi^R(\tau_g) \: = \: \chi^R(\tau_h \tau_g \tau_h^{-1} ).
\end{equation}
As a result, although characters of projective representations are not
invariant under conjugating group elements, they are invariant under
conjugating $\tau$'s.

Another important property of characters of projective representations is
that they vanish on non-$\omega$-regular group elements, see
e.g.~\cite[section 7.2, prop. 2.2]{karpilovsky}.

Now, we can define projectors, following \cite[section 7.3]{karpilovsky},
which are associated to irreducible projective representations,
and which form another basis for the center of the twisted
group algebra.  These are given by 
\begin{equation}  \label{eq:proj-defn}
P_R \: = \: \frac{\dim R}{|G|} \sum_{g \in G}
\frac{ \chi^R(g^{-1}) }{ \omega(g,g^{-1}) } \tau_g
\: = \:
 \frac{\dim R}{|G|} \sum_{g \in G} \chi^R(\tau_g^{-1}) \tau_g,
\end{equation}
where $R$ is an irreducible projective representation.
(Instead of summing over all group elements, one can equivalently
sum only over $\omega$-regular elements, as the character $\chi^R$ will
vanish on non-$\omega$-regular elements.)
These form a complete, mutually orthogonal, basis for the center of
the twisted group algebra, meaning that they obey
\begin{equation} \label{eq:proj-props}
P_R P_S \: = \: \delta_{RS} P_S, \: \: \:
\sum_R P_R \: = \: 1.
\end{equation}
They also obey the relation~(\ref{eq:delta-proj})
\begin{equation}
\delta(P_R) \: = \: \frac{ (\dim R)^2}{|G|}.
\end{equation}

These two bases (of twist fields, and of projectors) are related as follows:
\begin{equation} \label{eq:p-from-t}
P_R \: = \: \frac{\dim R}{|G|} \sum_{g \in G} 
\frac{ \chi^R\left( g^{-1} \right) }{ \omega(g,g^{-1})} \, T_{[g]},
\end{equation}
(which formally matches the result of taking the 
definition~(\ref{eq:proj-defn}) and replacing $\tau_g \in 
{\mathbb C}_{\omega}(G)$ with $T_{[g]}$, an element of the center),
and 
\begin{equation} \label{eq:t-from-p}
T_{[g]} \: = \: \sum_R \frac{ \chi^R(g) }{\dim R} \, P_R.
\end{equation}
These Fourier transforms are known, but for completeness,
as they are perhaps somewhat obscure,
next we will perform a consistency check and provide derivations.

As a consistency check, recall both $T_{[g]}$ and
$\chi^R(g)$ transform under conjugation.  However, using the
identity
\begin{equation}
\frac{ \omega(gh^{-1},h) }{ \omega(h,gh^{-1}) }
\: = \:
\frac{ \omega(h^{-1},hgh^{-1}) }{ \omega(g,h^{-1}) },
\end{equation}
a consequence of
\begin{equation}
(d\omega)(h^{-1},h,gh^{-1}) \,
(d\omega)(g,h^{-1},h) \: = \: 1,
\end{equation}
we see that both $T_{[g]}$ and $\chi^R(g)$ transform in the same way under
$g \mapsto hgh^{-1}$, and so the identity~(\ref{eq:t-from-p}) is
consistent.

As a consequence, if $C$ is any element of the center of the twisted
group algebra, it can be expressed similarly.  Write
\begin{equation}
C \: = \: \sum_{i=1}^n C_i T_{[h_i]},
\end{equation}
for $C_i \in {\mathbb C}$,
so that
\begin{equation}
\chi^R(C) \: = \: \sum_{i=1}^n C_i \chi^R(h_i),
\end{equation}
then from~(\ref{eq:t-from-p}) we have
\begin{eqnarray}
C & = &
\sum_{i=1}^m C_i \left[ \sum_R \frac{ \chi^R(T_{[h_i]}) }{ \dim R} P_R \right],
\\
& = &
\sum_R \frac{ \chi^R(C) }{\dim R} P_R.   \label{eq:c-from-p}
\end{eqnarray}

We can establish~(\ref{eq:p-from-t}) by direct computation, as follows.
\begin{eqnarray}
\lefteqn{
 \frac{\dim R}{|G|} \sum_{g \in G} 
\frac{ \chi^R\left( g^{-1} \right) }{ \omega(g,g^{-1})} \, T_{[g]}
} \nonumber \\
& = &
 \frac{\dim R}{|G|} \sum_{g \in G} 
\frac{ \chi^R\left( g^{-1} \right) }{ \omega(g,g^{-1})}
\frac{1}{|G|} \sum_h
\frac{ \omega(h,g)\, \omega(hg,h^{-1}) }{ \omega(h,h^{-1}) }
\tau_{hgh^{-1}},
\\
& = &
\frac{\dim R}{|G|^2} \sum_{h\in G} \frac{1}{\omega(h,h^{-1})} 
\tau_h \left[ \sum_{g \in G} \frac{ \chi^R(g^{-1}) }{ \omega(g,g^{-1}) }
\tau_g \right] \tau_{h^{-1}},
\\
& = &
\frac{1}{|G|} \sum_{h \in G} \frac{1}{\omega(h,h^{-1})} 
\, \tau_h \, P_R \, \tau_{h^{-1}},
\\
& = &
\frac{1}{|G|} \sum_{h \in G} \frac{\omega(h,h^{-1})}{\omega(h,h^{-1})} 
\, P_R,
\\
& = & P_R,
\end{eqnarray}
where we have used the fact that $P_R$ is central in the group algebra.

We can establish~(\ref{eq:t-from-p}) by direct computation, as follows.
\begin{eqnarray}
\sum_R \frac{ \chi^R(g) }{\dim R} \, P_R 
& = &
\sum_R \frac{ \chi^R(g) }{\dim R}
\frac{\dim R}{|G|} \sum_{h \in G} 
\frac{ \chi^R(h^{-1}) }{ \omega(h,h^{-1})} T_{[h]},
\\
& = &
\frac{1}{|G|} \sum_{h \in G} \left[
\sum_R \frac{ \chi^R(g) \chi^R(h^{-1}) }{ \omega(h,h^{-1}) }
\right] T_{[h]},
\\
& = &
\frac{1}{|G|} \sum_{h = a g a^{-1} \in [g]} 
\frac{|G|}{|[g]|}
\frac{ \omega(a,g) }{ \omega(h,a) }
T_{[aga^{-1}]},
\\
& = &
\frac{1}{|[g]|} \sum_{h = a g a^{-1} \in [g]} 
\frac{  \omega(a,g) }{ \omega(aga^{-1},a) }
\frac{ \omega(ga^{-1},a) }{ \omega(a,ga^{-1}) } T_{[g]},
\\
& = &
T_{[g]},
\end{eqnarray}
using the index formula~(\ref{eq:char-master2}) and the fact that
\begin{equation}
\frac{  \omega(a,g) }{ \omega(aga^{-1},a) }
\frac{ \omega(ga^{-1},a) }{ \omega(a,ga^{-1}) } 
\: = \: (d\omega)(a, ga^{-1},a) \: = \: 1.
\end{equation}

\subsection{Vacuum string amplitudes and  $\cH_0 \hookrightarrow \cH$  } 

We observe that the vacuum amplitudes of $G$-CTST are constructed by applying the delta-function on the twisted group algebra  ${\mathbb C}_{\omega}(G)$ to powers of a handle creation operator $\Pi$. We show in Section  \ref{sect:handle-op} that these powers generate a subspace of $\cZ ( {\mathbb C}_{\omega}(G)) $ with dimension equal to the number of distinct dimensions $(\dim R)$ of irreducible representations of ${\mathbb C}_{\omega}(G)$. In section \ref{sec:HandlesOnePt} we show that one point functions of twist fields on  higher genus surfaces can be used to determine sums of irreducible characters over irreducible representations having the same dimension. 

\subsubsection{The handle creation operator and twist fields}
\label{sect:handle-op}

One convenient way of expressing 
the partition function of (twisted) Dijkgraaf-Witten theory on a genus $h$
Riemann surface is as
\begin{equation}
Z_h \: = \: \frac{1}{|G|} \delta\left( \Pi^h \right),
\end{equation}
where $\Pi$ is the handle creation operator (a map 
${\mathbb C}_{\omega}(G) \rightarrow {\mathbb C}_{\omega}(G)$ which descends to
$\cH \rightarrow \cH$) which, for twisted theories,
is defined in section~\ref{sect:handlecreation}.

We can express the partition function more explicitly as follows.
Using the identity~(\ref{eq:handle-op-id}),
namely
\begin{equation}
\Pi \: = \: \sum_R \left( \frac{|G|}{\dim R} \right)^2 P_R ,
\end{equation}
so that
\begin{equation}
\Pi^h \: = \:  \sum_R \left( \frac{|G|}{\dim R} \right)^{2h} P_R
\end{equation}
(since $P_R$ is an idempotent),
and the identity~(\ref{eq:delta-proj}), namely
\begin{equation}
\delta(P_R) \: = \: \frac{ (\dim R)^2}{|G|},
\end{equation}
we have that the partition function is
\begin{eqnarray}
Z_h & = &
\frac{1}{|G|} \sum_R \left( \frac{|G|}{\dim R} \right)^{2h}
\delta(P_R), 
\\
& = &
|G|^{2h-2} \sum_R \left( \dim R \right)^{2-2h},
\end{eqnarray}
which matches the expression~(\ref{eq:partfn})  obtained independently. We conclude that 
\bea 
Z_{ h } = \frac{1}{|G|} \delta\left( \Pi^h \right) =  \sum_R \left( { |G| \over \dim R}  \right)^{2h-2}.
\eea
Using the formula \eqref{HandComb} for $\Pi$, the calculation of the delta function on the left-hand side can be done from the combinatorics of multiplying the elements $\tau_{ g } $ and picking up the coefficient of the identity. 
The formula, in the untwisted case, is well known in the mathematical literature \cite{Jones,Mednykh}. 
The combinatoric input from the left-hand side serves to give the power sums of $ { |G| \over \dim R} $. As explained in \cite{deMelloKoch:2021lqp}, we can go from the powers sums to the integers in a finite number of steps  by solving for the zeroes of a polynomial with integer coefficients. We further elaborate in section  \ref{sec:DetDiscon}  on the stringy interpretation of  the polynomial in the context of $G$-CTST.

Products of $ Z_h$, with appropriate symmetry factors, give us the vacuum sector of $G$-CTST. 
The vacuum sector of $G$-CTST defines two  distinguished subspaces of  
$\cH = \cZ ( \mC_{\omega} ( G ) ) $. Complex multiples of $\Pi$ form a one-dimensional subspace of $\cH$. 
Powers of $\Pi$ span a (generically)  higher-dimensional vector subspace of $\cH$.

\noindent 
{\bf  Proposition} The powers of the handle creation operator $\Pi$ 
span a vector subspace $\cH_0\hookrightarrow \cH$ which has dimension $D_0$ equal to the 
number of distinct integers $\dim R$ as $R$ runs over the set of irreducible projective
representations.

{\it Lemma (\cite[Lemma 2.1, Prop. 2.3]{DLS})} If we have a complete set of $L$ orthogonal
projectors $P_i$ acting on a vector space 
and take a linear combination with distinct coefficients $a_i$
\bea 
P \: = \: \sum_{ i=1 }^L a_i P_i,
\eea
then the powers of $P$ generate a space of dimension equal to $L$.  

{ \it Proof of proposition:} We can write 
\bea 
\Pi \:  = \: \sum_{ R } {  |G|^2 \over (\dim R)^2  }  P_R
\: = \: \sum_{ R' } 
{  |G|^2 \over (\dim R')^2  }   \tilde P_{ R' } 
\eea
where $R$ runs over all the distinct irreducible projective representations,
and $R'$ runs over a maximal list of irreducible projective representations
having distinct dimensions, while $\tilde P_{ R' }$ is a sum of the projectors for irreducible projective representations with the same dimension as $R'$.  The list of projectors $\tilde P_{ R'}$ spans a subspace 
$\cH_0 \hookrightarrow \cH$ of dimension $D_0$. In this subspace $\cH_0$, we can use the Lemma to show that the powers of $\Pi$ span $ \cH_0$.

The proposition has a physical interpretation in terms of the rank of a matrix of one-point functions in $G$-CTST.  
 Consider the one-point functions 
$M_{ l , [g] } \equiv \delta ( \Pi^l T_{ [g]} ) $, with $ l$ ranging from $1$ to $K$ and $g$ ranging over representatives of all the 
$\omega$-regular  conjugacy classes.  
(In the untwisted case this reduces to the set of all the conjugacy classes.)
This matrix has rank $D_0$.   
In the case where $M_{ l , [g] }$ is a  matrix with rational entries (this is the case for all untwisted cases and when the twists $\omega(g,h)$ can  all be chosen to be rational),   
an integer basis for the null space can be found using  discrete integer matrix algorithms   One approach is to use  algorithms for Hermite normal forms (such as algorithm 2.4.4 of \cite{HCohen}) 
and extract the null vectors  as explained for example in   
\cite[section 4.1]{JBSRKron}. Such discrete algorithms for null vectors are available in computational group theory software GAP \cite{gaponline}.  
This gives a combinatoric algorithm, 
starting from group multiplication combinatorics, 
which produces an interesting representation theoretic integer: the number of distinct $(\dim R)$ among the irreducible (projective) representations of a (twisted) group algebra.

\subsubsection{Character algorithm from higher genus one-point functions  }
\label{sec:HandlesOnePt}  

By considering the one-point functions
 $\delta ( \Pi^l T_{ [g]} ) $ on general genus, for fixed $[g]$, we can extract information about 
  characters of $\chi^R ( T_{\mu} ) /{\rm dim}\,R$.  Consider 
  \bea 
  { 1 \over |G| }  \delta ( \Pi^{ h }    T_{[g]}  ) 
& = & \sum_{ R  }  \left({  |G|^2 \over (\dim R)^2      } \right)^{h-1}  
 \frac{ \chi^R ( g ) }{\dim R} ,
\\
  & = & \sum_{ R'}\left({  |G|^2 \over (\dim R')^2     } \right)^{h-1} 
 \sum_{ R : R' }  { \frac{ \chi^R ( g ) }{ \dim R }  } ,
  \eea
  for the range $ l \in \{ 1, 2, \cdots , D_0 \}$,
where we have used the identity~(\ref{eq:final-handle-id2}). 
The primed sum runs over a maximal  set  $\{ R' \} $  of irreducible representations $R'$ having distinct dimensions.
The sum over $ \{ R : R' \}$  is a sum over the distinct irreducible
representations $R$ with the same dimension as $R'$. 
Let us define $\tilde R'$ to be the direct sum of irreducible projective
representations $R$ with the same dimension as $R'$. 
  Then we can write 
  \bea\label{OnePtChar} 
  { 1 \over |G| }  \delta \left( \Pi^{ h }    T_{ [g]}  \right)  \: =  \:
  \sum_{ R'} \left({  |G|^2 \over (\dim R')^2     } \right)^{h-1}
 { \chi^{ \tilde R'} ( g ) \over \dim  R' } . 
  \eea
  As $h$ runs over the set $\{ 1 , \cdots , D_0 \}$, we have a linear system of equations of size $D_0 \times D_0$ for the normalized characters  
${ \chi^{ \tilde  R'} ( g ) / \dim R' } $.    As $ R' $ and $ l $ range over the $D_0$ possibilities, we have 
a matrix 
\bea 
{ \mathcal V }_{ R' , h } =  \left({  |G|^2 \over (\dim R')^2     } \right)^{h-1}
\eea
of size $ D_0 \times D_0$. The equation \eqref{OnePtChar} takes the form 
\bea 
Y = { \mathcal V } \cdot X
\eea
where 
\bea 
&& Y_h  =  { 1 \over |G| }  \delta \left( \Pi^{ h }    T_{ [g]}  \right) , \cr 
&& X_{ R'}  =  { \chi^{ \tilde R'} ( g ) \over \dim  R' }  ,
\eea
and we recognize $ {\mathcal V  } $ as a Vandermonde matrix.
Since the $R'$ have been chosen to run over a  set of 
irreducible (projective) representations with distinct dimensions, the integers
$ \left ( {|G|^2 \over (\dim R')^2} \right ) $ are distinct. This ensures that ${\mathcal V } $ is invertible.
The inverse matrix can thus be used to construct the normalized characters $X_{ R'}$ from the combinatoric $G$-CTST data $Y_h$. As explained earlier, the construction of the ratios $\left({  |G|^2 \over (\dim R')^2     } \right ) $ from $G$-CTST data follows using the formulae in section \ref{sec:FTVac} in the twisted case, using the same algorithm described for the untwisted case in \cite{deMelloKoch:2021lqp}.

\section{Character algorithms and string amplitudes}\label{sec:CharBdy}

In the AdS/CFT correspondence, one is  led in connection with toy models of black hole information loss \cite{BCLS} 
 to consider questions of when sequences of central elements
suffice to distinguish representations and multiplicatively generate
the center of the group algebra \cite{KempRam}.   In the context of TQFTs such as
Dijkgraaf-Witten theory, it is natural to supplements such lists by the
handle creation operator.  To this end, in this section we present
some general statements about subsets that multiplicatively generate
the center of a (twisted) group algebra.
We also use these generating subspaces to give algorithms for the
construction of characters from string amplitudes in $G$-CTST. In the last subsection, we use these constructions to derive some integrality properties of characters and factorisation properties of character polynomials.

\subsection{Minimal generating subspaces of (twisted) group algebras }  
\label{subsecPismall}

We will say that a set of elements\footnote{
This argument expands the one presented in 
\cite{KempRam}.  It was described there for 
untwisted group algebras, but the extension to twisted group algebras which we develop here 
has the same form.  
} $\{ C_1, C_2, \cdots, C_k  \}$, with 
$C_i \in \cZ( \mC_{\omega} ( G) ) $, 
multiplicatively generate $\cZ ( \mC_{\omega} ( G ) ) $ 
if every element $T \in \cZ ( \mC_{\omega} ( G ) ) $ can be written as a 
linear combination of products of elements $C_i$: 
\bea 
T \: = \:  \sum_{ n_1 , n_2 , \cdots , n_k \ge 0 } t_{ n_1 , n_2 , \cdots , n_k } C_1^{n_1} C_2^{n_2} \cdots C_k^{n_k } . 
\eea
The coefficients $t_{ n_1, n_2 , \cdots , n_k  } $ are in $\mC $, and $C^{ 0 } $ is defined as $1$, the identity element of the group algebra.

\noindent 
{\bf Proposition} The following two statements are equivalent: \\
(1) A set of central elements $\{ C_1 , C_2, \cdots , C_k \}$ multiplicatively generate $ \cZ( \mC_{\omega} ( G )) $ \\
(2) The ordered lists of normalized characters $\{ { \chi^{R} ( C_1 ) \over \dim R }  , { \chi^R ( C_2 ) \over \dim R } \cdots , { \chi^R ( C_k ) \over \dim R }  \}$, for irreducible representations $R$ of $ \mC_{\omega} ( G ) $  distinguish the irreducible representations, 
i.e.~no two irreducible representations have the same list. 

The proof uses the fact that each element $C$ has an expansion in projectors $P_R$ given by~(\ref{eq:c-from-p}), which we repeat here: 
\bea 
C \: = \: \sum_{ R } { \chi^R ( C ) \over \dim R } P_R,
\eea
where the $P_R$ form a complete set of orthogonal projectors,
as in equation~(\ref{eq:proj-props}).
Consider first the case where $k =1$, and a single element $C_1 \in \cZ ( \mC_{\omega} ( G ) )  $ has the property that 
$\{ {\chi^R ( C_1 ) \over \dim R} \} $ distinguishes the irreducible representations $R$. The following fact is useful. 

{\bf Lemma } If $ T = \sum_{ R } a_R P_R $ with $a_R $ all distinct, then 
\bea\label{niceLem}  
P_R = \prod_{ S \ne R }  { ( T - a_S ) \over ( a_R - a_S ) }.  
\eea

We know that $\cZ ( \mC_{\omega} ( G ) ) $ is spanned by the projectors $P_R$. 
Since (in the case $k=1$) $C_1$ can be written as a linear combination of 
$P_R$ with distinct coefficients, the lemma above implies 
each $P_R$ can be written 
as a linear combination of powers of $C_1$, 
hence $\{ C_1 \} $ multiplicatively
generates $\cZ ( \mC_{\omega} ( G ) )$. 
The powers of $C_1$ range from $0$ up to a maximum of $K -1$ where $K$ is the dimension of $\cZ ( \mC_{\omega} ( G ) )$.

Suppose now that $k=2$, i.e. $\{ C_1, C_2 \}$ have lists of normalized characters  $\{ { \chi^{R} ( C_1) \over \dim R }, { \chi^R ( C_2 ) \over \dim R } \}$ 
which distinguish the irreducible representations $R$. We have 
\bea\label{C1RRp}  
C_1 \: = \: \sum_{ R } { \chi^R ( C_1 ) \over \dim R } P_R 
\: = \: \sum_{ R'} { \chi^{R'}  ( C_1 ) \over \dim R' } \widetilde { P_{R'}} ,
\eea
where $R'$ runs over a set  of irreducible representations with  distinct normalized characters 
$\chi^{R'}  ( C_1 ) / \dim {R'}$ and $\widetilde { P_{R'}}$ 
is the sum of projectors $P_R$ for all $R$ such that 
\bea 
{\chi^{R} ( C_1 ) \over \dim R } \: = \: { \chi^{R'} ( C_1 ) \over \dim {R'} } .
\eea

Let us define $[ C_1 ]_{ R' }$ to be this set of irreducible
representations $R$ with the same normalized characters as $R'$. 
Then we may write 
\bea 
\widetilde { P_{R'}} \: = \: \sum_{ R \in [ C_1 ]_{ R' } } P_R .
\eea
Let us denote the number of distinct $ R'$ in the sum for $C_1$ in  \eqref{C1RRp} by $K_1$, where
by assumption $K_1 \le K-1$. 
Using the Lemma, we can write each $\widetilde { P_{R'}}$ as a linear combination of powers of $C_1$. The largest power in these expressions is $ (K_1 -1)$. 
Consider now, for each $R'$, 
\bea 
\widetilde { P_{ R'} } C_2 \: = \:
\sum_{ R \in [ C_1]_{ R'} } { \chi^R ( C_2 ) \over \dim R } P_R .
\eea
By assumption, $\{ { \chi^{R} ( C_1) \over \dim R }  , { \chi^R ( C_2 ) \over \dim R } \}$ distinguish the irreducible representations, 
so it follows that for each $R'$, 
the ${ \chi^R ( C_2 ) \over \dim R } $  are distinct as
 $ R $ ranges over  the set $ [ C_1]_{ R'} $. 
This means that we can apply the Lemma to express 
 $P_R$ as a linear combination of  powers of the form  
\begin{equation}
 ( \widetilde { P_{ R'} } C_2 )^l \: = \: \widetilde { P_{ R'} } C_2^{ l }.
\end{equation}
Let $K_{a_1 ;  R'} $ be the number of elements $R$  in the set $[ C_1]_{ R'} $.
The powers $l$ range up to $ K_{a_1 ;  R'} -1 $.  
Since the $\widetilde { P_{ R'} } $ have already been expressed in terms of powers of $C_1$, we conclude that each $P_R$ can be expressed as a linear combination of powers of 
$\{ C_1 , C_2 \}$. 

We can express this more symmetrically by writing
\begin{eqnarray}
C_1 & = & \sum_{R_1' \in [C_1]} \frac{ \chi^{R_1'}(C_1) }{ \dim R_1'} 
\tilde{P}_{R_1'},
\\
C_2 & = & \sum_{R_2' \in [C_2]} \frac{ \chi^{R_2'}(C_2) }{ \dim R_2'} \tilde{P}_{R_2'}
\end{eqnarray}
where the sums run over representations with distinct normalized
characters, and the projectors $\tilde{P}$ are defined with respect
to the various sets $[C_i]$.

It is easy to see that this argument can be iterated for the cases of 
multiplicative generating subsets with more elements ($k > 2$).

We now describe another way to see that any projector $P_R $ is a linear combination of products of central elements $\{ C_1 , C_2 . \cdots , C_k \} $ with the property given in (2) of the proposition. For each $C_i$, we  can write 
\bea 
C_i \: = \: \sum_{ R } { \chi^R ( C_i ) \over \dim R } P_R 
\: = \: \sum_{ R_i'} { \chi^{R_i'}  ( C_i ) \over \dim R_i' } \widetilde { P_{R_i'}} ,
\eea
where $R_i'$ runs over a maximal set $S_i$ of irreducible
representations  with distinct normalized characters 
$ \chi^{R_i'}  ( C_i ) / \dim R_i'$. 
Let the cardinality of the set $S_i$ be $K_i$ and 
\bea 
\widetilde { P_{R_i'}} \: = \: \sum_{ R_i \in [ R_i' : C_i ] }  P_{ R_i } .
\eea
We have introduced the notation $[ R_i' : C_i ] $ for the set of irreducible
representations $R_i$ with the property that 
\bea
{ \chi^{ R_i  } ( C_i ) \over \dim R_i } \: = \:
 { \chi^{ R_i'  } ( C_i ) \over \dim R_i' } . 
\eea
The set $S_i$ is not unique because the sets $[ R_i' : C_i ] $ generically have more than one element, but we will make a choice of $S_i$. Using the Lemma, the projectors $\widetilde { P_{R_i'}}$ can be written as a linear combination of powers of $ C_i$. 
Now we know, by assumption,  that any irreducible representation $R$ 
is uniquely characterised by its normalised characters
\bea 
\left\{ { \chi^R ( C_1 ) \over \dim R } , { \chi^R ( C_2 ) \over \dim R }  , \cdots , { \chi^R ( C_k  ) \over \dim R }  \right\} .
\eea
This means that there is a unique list $ [ R_1' ( R ) , R_2'( R ) , \cdots , R_k' ( R ) ] $ with $R_1' (R) \in S_1, 
R_2'(R) \in S_2 , \cdots , R_k'( R ) \in S_k $, with the property that 
\bea 
\{ R  \} \: = \: [ R_1' (R)  ;  C_1 ]\cap [ R_2' ( R) ; C_2] \cap \cdots \cap [ R_k' (R)  ;   C_k ] .
\eea
This list is defined by the property that 
\begin{eqnarray}
{\chi^R ( C_1 ) \over \dim R } & = & {\chi^{R_1'(R) }  ( C_1 ) \over \dim R_1' },
\\
 { \chi^R ( C_2 ) \over \dim R } & = & {\chi^{R_2'(R) }  ( C_2 ) \over \dim R_2' },
\\
& \cdots  &
\\
 { \chi^R ( C_k ) \over \dim R } &  = & {\chi^{R_k' (R) }  ( C_2 ) \over \dim R_k' }.
\end{eqnarray}
It follows that 
\bea 
P_R  \: = \: \widetilde { P_{R_1' (R) }} \widetilde { P_{R_2' (R) }} \cdots \widetilde { P_{R_k' (R) }} .
\eea

In the next several subsections we will apply these ideas to examples of
sets of twist operators motivated by AdS/CFT, sometimes combined with
handle creation operators as also motivated by Dijkgraaf-Witten theory.

\subsubsection{Untwisted example:  ${\mathbb Z}_n$}
\label{sect:ex:zn:1}

The group ${\mathbb Z}_n$ has $n$ irreducible representations,
which we label $\rho_r$ for $r \in \{0, \cdots, n-1\}$.  If $g$ denotes
the generator of ${\mathbb Z}_n$, and $\xi = \exp(2 \pi i/n)$ the
generator of $n$th roots of unity, then
\begin{equation}
\rho_r(g) \: = \: \xi^r \: = \: \chi_r(g).
\end{equation}

From~(\ref{TpCp}), the twist fields are
\begin{equation}
T_{[g^k]} \: = \: \frac{1}{|G|} \sum_{h \in G} \tau_{hg^kh^{-1}}
\: = \: \tau_{g^k},
\end{equation}
and
from the definition~(\ref{eq:proj-defn}), we have that the projectors are
\begin{equation}
P_r \: = \: \frac{1}{n} \sum_{k=0}^{n-1} \chi_r(g^{-k}) \tau_{g^k}
\: = \: 
\frac{1}{n} \sum_{k=0}^{n-1} \xi^{-rk} \tau_{g^k}.
\end{equation}
In particular, in this case the center of the group algebra
${\mathbb C}({\mathbb Z}_n)$ coincides with the group algebra,
and has dimension $n$.

From~(\ref{eq:handle-op-id}) we have that the handle creation operator
is
\begin{eqnarray}
\Pi & = & \sum_{r=0}^{n-1} n^2 P_r,
\\
& = &
 \sum_{k=0}^{n-1} n^2 \left( \frac{1}{n}
\sum_{r=0}^{n-1} \xi^{-rk} \right) \tau_{g^k},
\\
& = & 
 \sum_{k=0}^{n-1} n^2 \delta_{k,0} \tau_{g^k},
\\
& = & 
n^2 \tau_1 \: = \: n^2.
\end{eqnarray}
Thus, we see that in this example
the handle creation operator and its powers can only ever
generate a one-dimensional subspace of the center of the group algebra. 
This is expected from section \ref{sec:FTVac}  
since all the irreducible representations
are one-dimensional, so the number of distinct values of 
$\dim R$ ($D_0$ in the discussion of  section \ref{sec:FTVac}) 
is $1$.

Now, let us turn to the question of constructing multiplicative
generators.  Consider for example the case of ${\mathbb Z}_3$.
Let $g$ denote the generator of the group, and $R_1$, $R_2$ the two
nontrivial representations, then the character table is given in
table~\ref{table:z3-chars},
where $\xi$ generates cube roots of unity.  In this case, we see that
the irreducible representations are uniquely determined by the
(normalized) characters of $g$, and it is also easy to check that
$T_{[g]}$ generates all the twist fields multiplicatively:
\begin{equation}
T_{[g]}^2 \: = \: T_{[g^2]},
\: \: \:
T_{[g]}^3 \: = \: 1 \: = \: T_{[1]}.
\end{equation}

\begin{table}[h]
\begin{center}
\begin{tabular}{c|ccc}
Representation & $1$ & $g$ & $g^2$ \\ \hline
$1$ & $1$ & $1$ & $1$ \\
$R_1$ & $1$ & $\xi$ & $\xi^2$ \\
$R_2$ & $1$ & $\xi^2$ & $\xi$
\end{tabular}
\end{center}
\caption{ \label{table:z3-chars}
Character table for ${\mathbb Z}_3$.
}
\end{table}

\subsubsection{Untwisted example:  $D_4$}
\label{sect:ex:d4:1}

List the elements of the dihedral group $D_4$ as
\begin{equation}
\{1, z, a, b, az, bz, ab, ba = abz\},
\end{equation}
where $z$ generates the ${\mathbb Z}_2$ center.

$D_4$ has five irreducible representations:  four one-dimensional
representations, and one two-dimensional representation.
The character table of $D_4$ is given in table~\ref{table:d4-chars}.

\begin{table}
\begin{center}
\begin{tabular}{c|crrrr}
& $\{1\}$ & $\{z\}$ & $\{a, az\}$ & $\{b, bz\}$ & $\{ab, ba\}$ \\ \hline
$1$ & $1$ & $1$ & $1$ & $1$ & $1$ \\
$1_a$ & $1$ & $1$ & $1$ & $-1$ & $-1$ \\
$1_b$ & $1$ & $1$ & $-1$ & $1$ & $-1$ \\
$1_{ab}$ & $1$ & $1$ & $-1$ & $-1$ & $1$ \\
$2$ & $2$ & $-2$ & $0$ & $0$ & $0$
\end{tabular}
\end{center}
\caption{ \label{table:d4-chars}
Character table of $D_4$ (without a twist).
}
\end{table}

Since there are five conjugacy classes (also five irreducible
representations), the center $\cZ( \mathbb{C} (D_4) )$ has dimension five.
Note, however, that knowing the normalized characters of just
two conjugacy classes suffices to distinguish characters.
For example, from table~\ref{table:d4-chars}, the characters of
$T_{[a]}$, $T_{[b]}$ suffice to distinguish all the irreducible
representations.  (By contrast, for example, the normalized
characters of $T_{[1]}$ and $T_{[z]}$ can
only be used to distinguish the two-dimensional representation from the
one-dimensional representation, but cannot distinguish between the
one-dimensional representations.)

This tells us that although the center $\cZ( \mC(D_4) )$ is a five-dimensional
vector space, it is generated multiplicatively by $T_{[a]}$ and
$T_{[b]}$, for example.
Indeed, 
from~(\ref{TpCp}) one finds
\begin{equation}
T_{[a]} \: = \: \frac{1}{2} \left( \tau_a \: + \: \tau_{az} \right),
\: \: \:
T_{[b]} \: = \: \frac{1}{2} \left( \tau_b \: + \: \tau_{bz} \right),
\end{equation}
and it is straightforward to check that
\begin{equation}
T_{[a]}^2 \: = \: \frac{1}{2} \left( 1 \: + \: \tau_z \right)
\: = \: T_{[b]}^2,
\: \: \:
T_{[a]} T_{[b]} \: = \: T_{[ab]},
\end{equation}
\begin{equation}
T_{[a]} (1 + \tau_z) \: = \: 2 T_{[a]},
\: \: \:
T_{[b]} (1 + \tau_z) \: = \: 2 T_{[b]}.
\end{equation}
Thus, the products of nonzero powers of $T_{[a]}$ and $T_{[b]}$ generate
themselves, $T_{[ab]}$, and the combination $1 + T_z$, and when we include
the zeroth power of $T_{[a]}$, $T_{[b]}$, we get all of the elements
of the center.

\subsubsection{Untwisted example: $S_n$}
\label{sect:ex:sn:1}

This question for the case of $S_n$ is motivated by AdS/CFT and was recently 
studied \cite{KempRam} in untwisted cases.
In that paper, central elements $T_{k}$ correspond to conjugacy classes
defined by permutations with a single non-trivial cycle of length $k$, and remaining cycles of length $1$. 
For any $\cZ ( \mC ( S_n) )$, the set $ \{ T_2 , T_3 , \cdots , T_n \}$ generates the center \cite{KempRam}  (Since there is no discrete torsion in this example,
twist fields depend only upon conjugacy classes, not upon representatives,
and so we only list the former.).

Typically a much smaller set $ \{ T_2 , T_3 , \cdots , T_{ k_* ( n ) } \} $ generates the center \cite{KempRam}, where 
$k_* (n)$ is much smaller than $n$, which is equivalent to the statement
that the normalized characters distinguish the irreducible representation $R$.
For example, the single normalized character 
$\chi_R ( T_2 ) /\dim R $ distinguishes $R$ for $n$ up to $5$ and $7$. The normalized characters of $T_2 , T_3 $ distinguish the Young diagrams up to $n=14$. Using the formulae for normalized characters given in \cite{Lasalle,CGS}
the lists $\{ { \chi_R ( T_2 ) \over \dim R } , { \chi_R ( T_3 ) \over \dim R }  \} $ were constructed for all the $R$ at fixed $n$, and verified (in Mathematica) to be distinct for $n$ up to $14$. For tests at higher $n$ (up to $80$) it was convenient to convert the question (using formulae in \cite{Lasalle,CGS}) of comparing lists of normalized characters to a question of comparing lists of power sums of contents of Young diagrams (for the precise procedure see \cite{KempRam}).

The discussion in \cite{KempRam} is generalised here to consider central elements including the handle creation operator, alongside the cycle operators. 
Using computations in GAP, we verify that the pairs  
\bea 
[ \dim R , \chi^R ( T_2 ) ]
\eea
uniquely determine all the Young diagrams of $S_n$ for $n $ up to $11$, as well as $13$. For example, the list of pairs 
at $n=6$ is: 
\bea
&& \{ [ 1, -15 ], [ 5, -45 ], [ 9, -45 ], [ 5, -15 ], [ 10, -30 ], [ 16, 0 ], 
  [ 5, 15 ], [ 10, 30 ], [ 9, 45 ], [ 5, 45 ], [ 1, 15 ]    \}.  
\nonumber
\eea
There is one such pair for every Young diagram. No two pairs are identical. 
Note that the list of ratios $\chi^R(T_2)/\dim R$ is 
 \bea 
 \{ [-15] , [ -9 ]  ,  [ -5 ] , [ -3 ] , [-3 ] , [ 0 ] , [3] , [ 3 ] , [ 5 ] , [ 9 ] , [15 ] \} .
 \eea
  These numbers are not unique: $-3$ and $3$ each appear twice. This means that $T_2$ does not generate the center of the group algebra of $S_6$ (as in \cite{KempRam})  but $\Pi $ and $T_2$ together do. 
  
 Computations in GAP also show the lists $\{ \dim R , \chi^R ( T_2 ) , \chi^R ( T_3 ) , \chi^R ( T_4 ) \} $ distinguish all the 
irreducible representations for $S_n$ at $n $ up to at least $30$. This means that the center is generated by $ \{ \Pi , T_2 , T_3 , T_4 \}$ for $\mC ( S_n) $ with $n $ up to at least $30$.

For later comparisons, we give the character table of
$S_4$ in table~\ref{table:s4-chars}, from \cite[table 4.5]{hh}.

\begin{table}[h]
\begin{center}
\begin{tabular}{c|rrrrr}
Irrep & $(1^4)$ & $(2 1^2)$ & $(2^2)$ & $(31)$ & $(4)$ \\ \hline
$1$ & $1$ & $1$ & $1$ & $1$ & $1$ 
\\
$R_2$ & $1$ & $-1$ & $1$ & $1$ & $-1$ 
\\
$R_3$ & $2$ & $0$ & $2$ & $-1$ & $0$ 
\\
$R_4$ & $3$ & $1$ & $-1$ & $0$ & $-1$ 
\\
$R_5$ & $3$ & $-1$ & $-1$ & $0$ & $1$
\end{tabular}
\end{center}
\caption{ \label{table:s4-chars}
Character table of $S_4$, from \cite[table 4.5]{hh}.
Conjugacy classes are indicated by the number of elements exchanged.
For example, a ``$1$'' indicates that an element is mapped to itself,
whereas a ``$4$'' indicates that all four elements are permuted,
for example $1 \mapsto 2 \mapsto 3 \mapsto 4$.
(The fact that this distinguishes conjugacy classes is discussed in
e.g.~\cite[theorem 3.7]{hh}.)
In particular, $(1^4)$ is the conjugacy class of the identity.
}
\end{table}

\subsubsection{Untwisted example: $\tilde{S}_n$}
\label{sect:ex:untwisted-tildes}

The group $\tilde{S}_n$ is a central extension of the symmetric
group $S_n$ by ${\mathbb Z}_2$:
\begin{equation}
1 \: \longrightarrow \: {\mathbb Z}_2 \: \longrightarrow \:
\tilde{S}_n \: \longrightarrow \: S_n \: \longrightarrow \: 1.
\end{equation}
It is described in \cite[chapter 2]{hh}
by generators $z, t_1, t_2, \cdots, t_{n-1}$ and relations
\begin{equation}
z^2 \: = \: 1, 
\: \: \:
z t_j \: = \: t_j z,
\: \: \:
t_j^2 \: = \: z,
\end{equation}
\begin{equation}
(t_j t_{j+1})^2 \: = \: z \: \: \: \mbox{ for } 1 \leq j \leq n-2,
\end{equation}
\begin{equation}
t_j t_k \: = \: z t_k t_j \: \: \: \mbox{ for } |j-k| > 1
\mbox{ and } 1 \leq j, k \leq n-1.
\end{equation}

The character table of $\tilde{S}_4$ is given in table~\ref{table:ts4-chars}
(from \cite[table 4.7]{hh}).

\begin{table}[h]
\begin{center}
\begin{tabular}{c|rrrrrrrr}
Irrep & $(1^4)'$ & $(1^4)''$ & $(2 1^2)$ & $(2^2)$ & $(31)'$ & $(31)''$ &
$(4)'$ & $(4)''$ \\ \hline
$1$ & $1$ & $1$ & $1$ & $1$ & $1$ & $1$ & $1$ & $1$ \\
$R_2$ & $1$ & $1$ & $-1$ & $1$ & $1$ & $1$ & $-1$ & $-1$ \\
$R_3$ & $2$ & $2$ & $0$ & $2$ & $-1$ & $-1$ & $0$ & $0$ \\
$R_4$ & $3$ & $3$ & $1$ & $-1$ & $0$ & $0$ & $-1$ & $-1$
\\
$R_5$ & $3$ & $3$ & $-1$ & $-1$ & $0$ & $0$ & $1$ & $1$ 
\\
$R_6$ & $2$ & $-2$ & $0$ & $0$ & $1$ & $-1$ & $\sqrt{2}$ & $-\sqrt{2}$ 
\\
$R_7$ & $2$ & $-2$ & $0$ & $0$ & $1$ & $-1$ & $-\sqrt{2}$ & $\sqrt{2}$ 
\\
$R_8$ & $4$ & $-4$ & $0$ & $0$ & $-1$ & $1$ & $0$ & $0$
\end{tabular}
\end{center}
\caption{ \label{table:ts4-chars}
Character table for $\tilde{S}_4$, from \cite[table 4.7]{hh}.
The notation for conjugacy classes references their images in $S_4$,
which
are indicated with the same notation
as in table~\ref{table:s4-chars}.  The primes refer to differences
arising from including the central element $z$.
For example, $(1^4)'$ is the conjugacy class of the identity,
whereas $(1^4)''$ is the conjugacy class of $z$.
See \cite[theorem 3.8]{hh} for further details.
}
\end{table}

From table~\ref{table:ts4-chars}, we see for example that the normalized
characters of the conjugacy classes $(31)'$ and $(4)'$ uniquely
distinguish all the representations, hence, using the proposition in section \ref{subsecPismall} ,  we expect that the 
center $\cZ( \mC(\tilde{S}_4))$ is multiplicatively generated by
twist fields corresponding to those two elements.

\subsubsection{Twisted example: ${\mathbb Z}_2 \times {\mathbb Z}_2$}
\label{sect:twisted:z2z2:1}

Let us now turn to a simple twisted example,
namely $G = {\mathbb Z}_2 \times {\mathbb Z}_2$,
with $[\omega] \in H^2(G, U(1))$ the nontrivial element.
A representative 2-cocycle $\omega$ is
\begin{eqnarray}
\omega(a,b) \: = \: \omega(b,ab) \: = \: \omega(ab,a) & = & +i,
\\
\omega(b,a) \: = \: \omega(ab,b) \: = \: \omega(a,ab) & = & -i,
\end{eqnarray}
where $G = \langle a, b \rangle$, and with $\omega(g,h) = +1$ for other
$g, h$.  

The only $\omega$-regular conjugacy class in this case is $\{1\}$.
From the definition~(\ref{TpCp}), the twist fields are
\begin{equation}
T_{[1]} \: = \: \tau_1 \: = \: 1, \: \: \:
T_{[a]} \: = \: 0 \: = \: T_{[b]} \: = \: T_{[ab]}.
\end{equation}
(Although there is only one $\omega$-regular conjugacy class,
we can certainly compute twist fields for other conjugacy classes,
though as we see we do not get any further twist fields.)

There is only one irreducible projective representation
\cite[section 3.7]{karpilovsky}, which we label
$\rho$.  It is two-dimensional, and for the
2-cocycle above can be represented by
\begin{equation}
\rho(1) \: = \: \left[ \begin{array}{rr} 1 & 0 \\ 0 & 1 \end{array} \right],
\: \: \:
\rho(a) \: = \: \left[ \begin{array}{rr} 0 & 1 \\ 1 & 0 \end{array} \right],
\: \: \:
\rho(b) \: = \: \left[ \begin{array}{rr} 0 & -i \\ i & 0 \end{array} \right],
\: \: \:
\rho(ab) \: = \: \left[ \begin{array}{rr} 1 & 0 \\ 0 & -1 \end{array} \right],
\end{equation}
in the sense that
\begin{equation}
\rho(g) \rho(h) \: = \: \omega(g,h) \, \rho(gh).
\end{equation}

From this and the definition~(\ref{eq:proj-defn}), one quickly
computes that the single projector is given by
\begin{equation}
P_{\rho}
\: = \: \frac{\dim \rho}{|G|} \sum_{g \in G} 
\frac{ \chi^{\rho}\left( g^{-1} \right) }{ \omega(g,g^{-1}) }
\tau_g
 \: = \: \tau_1 \: = \: 1,
\end{equation}
essentially because only $\rho(1)$ has a nonzero trace.
Then, using the identity~(\ref{eq:handle-op-id}), the handle creation operator
is easily computed to be
\begin{equation}
\Pi \: = \: \sum_R \left( \frac{ |G| }{ \dim R } \right)^2 P_R
\: = \: 4 P_{\rho} \: = \: 4.
\end{equation}

In this case, the center of the twisted group algebra is also 
one-dimensional, corresponding to complex multiples of the identity,
and so $\Pi$ generates the center, essentially trivially.

In passing, let us also compare to the character table of $D_4$,
table~\ref{table:d4-chars}.
Since $D_4$ is an extension of ${\mathbb Z}_2 \times {\mathbb Z}_2$,
it includes information about the irreducible projective representation of
${\mathbb Z}_2 \times {\mathbb Z}_2$, which in this case is an
honest representation of $D_4$.
Looking at table~\ref{table:d4-chars}, we see the first four $D_4$
representations descend to representations of ${\mathbb Z}_2 \times
{\mathbb Z}_2$, because they take the same value on $z$ as on the identity.
The fifth representation, the two-dimensional one, takes a different
value on $z$ than on $1$, and so does not arise from an ordinary
representation of ${\mathbb Z}_2 \times {\mathbb Z}_2$.  This representation
corresponds to the irreducible projective representation of
${\mathbb Z}_2 \times {\mathbb Z}_2$.

\subsubsection{Twisted example:  $D_4$}
\label{sect:twist:d4:1}

Now, consider the $2n$-element dihedral group $G = D_n$.
This can be generated by $a$, $b$, such that
\begin{equation}
a^2 \: = \: 1, \: \: \: b^n \: = \: 1, \: \: \:
a b a \: = \: b^{-1}.
\end{equation}
For simplicity, we assume $n$ is even.
This has a nontrivial element of $H^2(D_n,U(1))$, given by
\begin{equation}
\omega(b^i, b^j a^k) \: = \: 1, \: \: \:
\omega(b^i a, b^j a^k) \: = \: \epsilon^j,
\end{equation}
where $\epsilon$ generates the $n$th roots of unity.
For $n$ even, $b^{n/2}$ is central,
and the dihedral group $D_n$ has $n/2$ irreducible projective
representations, each two-dimensional, described as follows\footnote{
In our conventions we exchanged the roles of $a$ and $b$ relative
to \cite[section 3.7]{karpilovsky}.
}
\cite[section 3.7]{karpilovsky}.  For
$r \in \{1, \cdots, n/2\}$, define
\begin{equation}
A_r \: = \: \left[ \begin{array}{cc} 0 & 1 \\ 1 & 0 \end{array} \right],
\: \: \:
B_r \: = \: \left[ \begin{array}{cc} \epsilon^r & 0 \\ 0 & \epsilon^{1-r}
\end{array} \right],
\end{equation}
and then the $r$th representation is given by
\begin{equation}
\rho_r(b^i a^j) = B_r^i A_r^j,
\end{equation}
for $i \in \{0, \cdots, n-1\}$ and $j \in \{0, 1\}$.

To make this more concrete, we specialize to $D_4$, which has center
${\mathbb Z}_2$, generated by $b^2$.  Here,
$H^2(D_4,U(1)) = {\mathbb Z}_2$, with a representative of the nontrivial cocycle
given above.
The conjugacy classes in $D_4$ are
\begin{equation}
\{1 \}, \: \: \: 
\{b^2\}, \: \: \:
\{b, b^3\}, \: \: \:
\{a, ab^3\}, \: \: \:
\{ab, ab^3\},
\end{equation}
of which only two are $\omega$-regular, namely $\{1\}$ and $\{b, b^3\}$.
From~(\ref{TpCp}), twist fields are
\begin{equation}
T_{[1]} \: = \: \tau_1 \: = \: 1, \: \: \:
T_{[b]} \: = \: \frac{1}{2}\left( \tau_b + \epsilon \tau_{b^3} \right),
\: \: \:
T_{[b^3]} \: = \: \frac{1}{2}\left( \epsilon^3 \tau_b + \tau_{b^3} \right),
\end{equation}
\begin{equation}
T_{[b^2]} \: = \: 0 \: = \: T_{[a]} \: = \: T_{[ab]},
\end{equation}
where $\epsilon$ generates fourth roots of unity, hence we can
take $\epsilon = i$.
(Only for the $\omega$-regular conjugacy classes are the twist fields
produced by~(\ref{TpCp}) nonzero.  Also, although $b$, $bz$ are in the
same equivalence class, $T_{[g]}$ is not invariant under conjugation,
but instead are related by~(\ref{eq:twist-conj}), as is easily checked
to relate $T_{[b]}$, $T_{[b^3]}$ above.)

Since there are two $\omega$-regular conjugacy classes, 
there are
two (two-dimensional) irreducible
projective representations, which are given by
\begin{equation}
\rho_r(1) \: = \: I, \: \: \:
\rho_r(a) \: = \: \left[ \begin{array}{cc} 0 & 1 \\ 1 & 0 \end{array} \right],
\: \: \:
\rho_r(b) \: = \: \left[ \begin{array}{cc} \epsilon^r & 0 \\
0 & \epsilon^{1-r} \end{array} \right],
\end{equation}
\begin{equation}
\rho_r(b^2) \: = \: \left[ \begin{array}{cc} \epsilon^{2r} & 0 \\
0 & \epsilon^{2-2r} \end{array} \right],
\: \: \:
\rho_r(b^3) \: = \: \left[ \begin{array}{cc} \epsilon^{3r} & 0 \\
0 & \epsilon^{3-3r} \end{array} \right],
\: \: \:
\rho_r(ba) \: = \: \left[ \begin{array}{cc} 0 & \epsilon^4 \\
\epsilon^{1-r} & 0 \end{array} \right],
\end{equation}
\begin{equation}
\rho_r(b^2 a) \: = \: \left[ \begin{array}{cc} 0 & \epsilon^{2r} \\
\epsilon^{2-2r} & 0 \end{array} \right],
\: \: \:
\rho_r(b^3 a) \: = \: \left[ \begin{array}{cc} 0 & \epsilon^{3r} \\
\epsilon^{3-3r} & 0 \end{array} \right].
\end{equation}
Since there are two irreducible projective representations,
the twisted group algebra of $D_4$ has a two-dimensional center.
We give the character table for projective representations of $D_4$ in
table~\ref{table:td4-chars}.

\begin{table}[t]
\begin{center}
\begin{tabular}{c|cccccc}
$r$ & $1$ & $b^2$ & $b$ & $b^3$ & $\{a, ab^2\}$ & $\{ab, ab^3\}$ \\ \hline
$r=1$ & $2$ & $0$ & $+(1+i)$ & $+(1-i)$ & $0$ & $0$ \\
$r=2$ & $2$ & $0$ & $-(1+i)$ & $-(1-i)$ & $0$ & $0$
\end{tabular}
\end{center}
\caption{\label{table:td4-chars}
Character table for irreducible projective representations of $D_4$.
Note that although $b$ and $b^3$ are in the same conjugacy class of $D_4$,
their characters are different, so we list them separately.
Also note that only characters of representatives of $\omega$-regular
conjugacy classes are nonzero.
}
\end{table}

Plugging into~(\ref{eq:proj-defn}), we have
\begin{equation}
P_r \: = \: \frac{1}{4} \left[ 2 \: + \:
(\epsilon^{3r} + \epsilon^{3-3r}) \tau_b \: + \:
(\epsilon^r + \epsilon^{1-r}) \tau_{b^3} \right],
\end{equation}
using the fact that $\epsilon^{2r} + \epsilon^{2-2r} = 0$.
(As a consistency check, it is straightforward to show that
$P_r^2 = P_r$, $P_1 P_2 = 0$, and $P_1 + P_2 = 1$.)

From~(\ref{eq:handle-op-id}), we have
\begin{equation}
\Pi \: = \: \sum_r \left( \frac{ |D_4| }{ \dim \rho_r } \right)^2 P_r
\: = \: (16)(P_1 + P_2) \: = \: 16,
\end{equation} 
using the fact that $P_1 + P_2 = 1$.
We see immediately that $\Pi^2 \propto \Pi$, and so
the handle creation operator generates a one-dimensional subspace
of the two-dimensional center of the twisted group algebra of $D_4$.
On the other hand, note that
\begin{equation}
T_{[b]}^2 \: = \: \epsilon/2 \: \propto \: 1,
\end{equation}
hence the center can be multiplicatively generated by $T_{[b]}$ alone,
which is consistent with table~\ref{table:td4-chars}.

\subsection{Character algorithms and generating subspaces }\label{sec:char-gen} 

In \cite[section 3.1]{deMelloKoch:2021lqp} the first author and his collaborators
interpreted the Burnside construction \cite{Burnside}( see \cite{Dixon,Schneider} for subsequent improvements) 
in terms of (untwisted)
combinatoric amplitudes 
on genus one surfaces.
The key formula, which takes the same form in the twisted case,
is~(\ref{eq:final-handle-id2}), which implies
\bea\label{charbdyBurnside}
 { 1 \over |G| } \delta ( \Pi T_{[g]}^l  ) \: = \: 
\sum_{ R }\left ( { \chi^R ( g ) \over \dim R  }   \right )^l .
\eea
Using the power sums, we solve a polynomial equation to get 
the normalized characters for the twist fields $T_{[g]}$.
The polynomial equation is actually the eigenvalue equation for the 
matrix of structure constants 
$({\mathfrak C}_{[g]})_{\alpha }^{ \beta } = {\mathfrak C}_{[g] \alpha }^{ \beta }$ 
where $T_{[g]} T_{ \alpha } = {\mathfrak C}_{ [g] \alpha }^{ \beta } T_{ \beta }$. 
After the normalized characters have been found, 
the dimensions can be found using the orthogonality 
relation~(\ref{eq:char-master2}), which implies 
\bea 
\sum_{R} {1 \over \omega(g,g^{-1})} 
 \chi^R( g )  \chi^R ( g^{-1} )  
 \:  = \:  { |G| \over | [g] | }.
\eea

It is interesting to consider the implications for the  character algorithms 
of knowing a subset of ($\omega$-regular)
conjugacy classes whose normalized characters determine the irreducible representations. 
Suppose a set of central elements
$\{ C_1 , C_2 , \cdots , C_k \}$ (possibly including
$\Pi$)  are known to 
multiplicatively generate the center $\cZ ( \mC_{ \omega } ( G ) )$
of a (possibly twisted) 
group algebra. 
In the case of the untwisted group algebra of $S_n$ (for $n < 80 $)  it has been shown \cite{KempRam}  that there are interesting small (compared to $n$) subsets which have this property. 
In section~\ref{sec:fin-min-gen} we explain how to find such minimal generating subsets.

Let us first consider the case where a single operator 
$C_1 \in \cZ( \mC_{\omega}(G) )$
multiplicatively generates the center, as we have seen occurs in
examples in sections~\ref{sect:ex:zn:1}, \ref{sect:ex:sn:1}. 
In this case, following a construction similar to the use of the Vandermonde matrices in section \ref{sec:FTVac}, 
we can compute the characters of any (represented, $\omega$-regular)
conjugacy class $\cC_{ \mu}$ from the genus one amplitudes 
associated with $C_1, T_{\mu}$ and the normalized characters of
$C_1$. 
Specifically we start with the string amplitudes~(\ref{eq:final-handle-id2})
\begin{eqnarray}\label{genTaChar}
{ 1 \over |G| } \delta ( \Pi C_1^k T_{ \mu } ) 
& =  & 
\sum_{ R } \left ( { \chi^R ( C_1 ) \over \dim R  }   \right )^{ k } { \chi^R ( T_{\mu  } ) \over \dim R } .
\end{eqnarray}
(It suffices to only consider
$k \in \{ 0 , 1, \cdots , K -1 \}$, 
where $K$ is the number of conjugacy classes.)

In terms of the Vandermonde matrix
\bea 
\mathcal{ V }_{ k , R } \: = \:
 \left ( { \chi^R ( C_1 ) \over \dim R  }   \right )^{ k }
\eea
the expression~(\ref{genTaChar}) is an invertible linear system of 
equations relating string amplitudes to the 
normalized characters of $T_{ \mu}$. 
By using the inverse of the Vandermonde matrix, we can solve for the 
normalized characters $\chi^R ( T_{\mu  } ) / \dim R$ in terms of the 
string amplitudes in \eqref{genTaChar} and the normalized characters
of the generator $C_1$, both assumed known.

Suppose now that $\{ C_1 , C_2 \}$ are a minimal set that
multiplicatively
generate
$\cZ ( \mC_{\omega} ( G ) )$,
as we have seen in examples in
sections~\ref{sect:ex:d4:1}, \ref{sect:ex:sn:1}, \ref{sect:twist:d4:1}. 
In such cases the lists 
$\{   { \chi^{R } ( C_1 ) / \dim R } , { \chi^R ( C_2 ) / \dim R  } \}$ 
uniquely determine the irreducible representations $R$. 
Now, 
we can again consider the problem of determining the normalized characters 
for a general conjugacy class (with specified representative)
$T_{\mu}$, from the string amplitudes. Start with the amplitudes  
\bea 
{ 1 \over |G| } \delta ( \Pi C_1^k T_{ \mu } ) \: 
= \: \sum_{ R } \left ( { \chi^R ( C_1 ) \over \dim R  }   \right )^{ k } { \chi^R ( T_{\mu  } ) \over \dim R  }.
\eea
(As before, it suffices to restrict to
$k \in \{ 0, 1,  \cdots , K' - 1 \} $, 
where $K'$ is the number of distinct normalized characters 
$ { \chi_R ( C_1 ) / \dim R  } $.) Let $R'$ run over a set of irreducible representations (of size $K'$) with distinct normalized characters
$ { \chi_{R'}  ( C_1 ) / \dim R'  } $,
and $[R:R']$ over the irreducible representations with the same normalized characters as
$R'$. We write 
 \bea 
 { 1 \over |G| } \delta ( \Pi C_1^k T_{ \mu } ) 
= \sum_{ R' } \left ( { \chi^{R'} ( C_1 ) \over \dim R'  }   \right )^{ k } 
\sum_{ [ R : R'] } { \chi^{R} ( T_{\mu  } ) \over \dim R  }.
 \eea
By inverting\footnote{
The reader should note that if for example $C_1 = T_{[1]} = 1$, then the
Vandermonde matrix may not be invertible, and this procedure would not work.
However, we exclude that case from consideration, by restricting to
minimal generating sets.
} the $ K' \times K'$  Vandermonde matrix with matrix elements 
\bea\label{VDKprime} 
{\mathcal V}_{k,R'} \: = \:
\left ( { \chi^{R'} ( C_1 ) \over \dim R'  }   \right )^{ k } 
\eea
we now determine the sums 
\bea 
\sum_{ [ R : R'] } { \chi^{R} ( T_{\mu  } ) \over \dim R  }
\eea
ranging over the distinct irreducible representations $R$ having the same normalized character 
${ \chi_{ R } ( C_1 ) / \dim R } $  as $R'$. 
We denote the  number of such $R$ (the number of elements
of $[R:R']$) by $D_{ 1 ; R' } $.

Using the fact that 
\begin{equation}
\left\{ { \chi_R ( C_1 ) \over \dim R }  , { \chi_R ( C_2 ) \over \dim R }  
\right\}
\end{equation}  
distinguish all irreducible representations, we know that for any $R'$, as $R$ ranges over the set $ [ R : R'] $, 
the list $\{ {\chi_{ R } ( C_2 ) / \dim R }\} $ has no repeated elements. Now for each $R'$, and each 
 $l \in \{ 0 , 1, \cdots , D_{ 1 , R'} -1  \}$ we can consider 
\bea\label{C1C2sys}  
{ 1 \over |G| } \delta \left ( \Pi C_1^{ k }  ( C_2^l T_{ \mu } )  \right ) 
&& = \sum_{ R }\left( { \chi^R ( C_1 ) \over \dim R }  \right )^{ k  }  \left( { \chi^R ( C_2 ) \over \dim R } \right)^{ l } { \chi^R ( T_{\mu  } ) \over \dim R }, \cr
&& = \sum_{ R' } \left ( { \chi^{R'} ( C_1 ) \over \dim R'  }   \right )^{ k } 
\sum_{ [ R : R'] }  \left( { \chi^R ( C_2 ) \over \dim R } \right)^{ l } { \chi^{R} ( T_{\mu  } ) \over \dim R  }.
\eea
As $ k $ ranges over $ \{ 0, 1, \cdots , K' -1 \}$, we have a 
linear system  of equations for 
\begin{equation}
\sum_{ { R : R'}} \left( { \chi^R ( C_2 ) \over \dim R } \right)^{ l } { \chi^{R} ( T_{\mu  } ) \over \dim R  } 
\end{equation}
given by the invertible $ K' \times K'$  Vandermonde matrix \eqref{VDKprime}. 
By using the inverse of the Vandermonde matrix, we obtain 
\bea 
\sum_{ { R : R'} }\left( { \chi^R ( C_2 ) \over \dim R } \right)^{ l } { \chi^{R} ( T_{\mu  } ) \over \dim R  }.
\eea
Collecting the results for all the $l \in \{ 0 , 1, \cdots , D_{ 1, R' } -1 \}$, we now have a linear system for 
${ \chi^{R} ( T_{\mu  } ) / \dim R  } $ 
for all the $R$  in the set $[ R ; R']$, 
given by the invertible $ D_{ 1 , R'} \times D_{ 1 , R'} $ Vandermonde matrix with matrix elements 
\bea 
{\mathcal V}_{\ell,R} \: = \:
\left( { \chi^R ( C_2 ) \over \dim R } \right)^{ l }.
\eea
By inverting the Vandermonde matrix, we obtain ${ \chi^{R} ( T_{\mu  } ) / \dim R  }$ for all $R$ with the property that 
\begin{equation}
{ \chi^R ( C_1 ) \over \dim R } \: = \:{  \chi^{R'}  ( C_1 ) \over \dim R' }.
\end{equation}

It is clear that the above procedure can be iterated to give a procedure for constructing normalized characters $T_{ \mu}$ in cases where a longer list 
\begin{equation}
\left\{ { \chi^R ( C_1 ) \over \dim R } , { \chi^R ( C_2 ) \over \dim R } , \cdots , { \chi^R ( C_k ) \over \dim R }\right\}  
\end{equation}
distinguish the irreducible representations (equivalently $\{ C_1 , C_2 , \cdots , C_k \}$ generate the center). Note that the generating set of central elements can all be obtained by averaging over fixed conjugacy  classes, and may also include central operators such as the handle operator $\Pi$ as discussed in section~\ref{subsecPismall}.

\subsubsection{Untwisted example:  ${\mathbb Z}_n$}

In this section we will illustrate the method in a case with well-known
results, specifically, the case
$G = {\mathbb Z}_3$.  

As discussed in section~\ref{sect:ex:zn:1}, if $g$ generates the
group ${\mathbb Z}_n$, then $T_g$ generates the center multiplicatively.
Following the prescription given above, the Dijkgraaf-Witten amplitudes
determine the normalized characters of any other conjugacy class.
Specifically, write
\begin{equation}
\frac{1}{|G|} \delta\left( \Pi T_{[g]}^k T_{\mu} \right)
\: = \: \sum_R {\mathcal V}_{k,R} \frac{ \chi^R(T_{\mu}) }{ \dim R},
\end{equation}
where
\begin{equation}
{\mathcal V}_{k,R} \: = \: \left( \frac{ \chi^R(T_{[g]}) }{ \dim R} \right)^k.
\end{equation}
Using table~\ref{table:z3-chars},
we have
\begin{eqnarray}
{\mathcal V}_{k,R=1} & = &
1,
\\
{\mathcal V}_{k,R=R_1} & = &
\xi^k,
\\
{\mathcal V}_{k,R=R_2} & = &
\xi^{2k},
\end{eqnarray}
for $\xi$ a generator of cube roots of unity, hence
\begin{equation}
{\mathcal V} \: = \: \left[ \begin{array}{ccc}
1 & 1 & 1 \\
1 & \xi & \xi^2 \\
1 & \xi^2 & \xi \end{array} \right],
\: \: \:
{\mathcal V}^{-1} \: = \: \frac{1}{3} \left[ \begin{array}{ccc}
1 & 1 & 1 \\
1 & \xi^2 & \xi \\
1 & \xi & \xi^2 \end{array} \right].
\end{equation}

Let us also take as given the string amplitudes
\begin{eqnarray}
\frac{1}{|G|} \delta\left( \Pi T_{[g]}^k T_{1} \right) 
& = &
3 \delta_{0,k \mod 3},
\\
\frac{1}{|G|} \delta\left( \Pi T_{[g]}^k T_{[g]} \right) 
& = &
3  \delta_{0,k+1 \mod 3},
\\
\frac{1}{|G|} \delta\left( \Pi T_{[g]}^k T_{[g^2]} \right) 
& = &
3  \delta_{0,k+2 \mod 3}.
\end{eqnarray}

From these string amplitudes we then compute
\begin{equation}
\left( \frac{\chi^R(T_{[1]}) }{ \dim R} \right) \: = \:
{\mathcal V}^{-1} \left[ \begin{array}{c} 3 \\ 0 \\ 0 \end{array}
\right]
\: = \: \left[ \begin{array}{c} 1 \\ 1 \\ 1 \end{array} \right],
\end{equation}
matching the known result
\begin{equation}
\frac{ \chi^R(T_{[1]}) }{ \dim R} \: = \: 1
\end{equation}
for each representation $R$.  Similarly,
\begin{equation}
\left( \frac{\chi^R(T_{[g]}) }{ \dim R} \right) \: = \:
{\mathcal V}^{-1} \left[ \begin{array}{c} 0 \\ 0 \\ 3 \end{array}
\right]
\: = \: \left[ \begin{array}{c} 1 \\ \xi \\ \xi^2 \end{array} \right],
\end{equation}
matching the result
\begin{equation}
\frac{\chi^1(T_{[g]})}{\dim 1} \: = \: 1, \: \: \:
\frac{\chi^{R_1}(T_{[g]})}{\dim R_1} \: = \: \xi, \: \: \:
\frac{\chi^{R_2}(T_{[g]})}{\dim R_2} \: = \: \xi^2.
\end{equation}
Finally,
\begin{equation}
\left( \frac{\chi^R(T_{[g^2]}) }{ \dim R} \right) \: = \:
{\mathcal V}^{-1} \left[ \begin{array}{c} 0 \\ 3 \\ 0 \end{array}
\right]
\: = \: \left[ \begin{array}{c} 1 \\ \xi^2 \\ \xi \end{array} \right],
\end{equation}
matching the result
\begin{equation}
\frac{\chi^1(T_{[g]})}{\dim 1} \: = \: 1, \: \: \:
\frac{\chi^{R_1}(T_{[g]})}{\dim R_1} \: = \: \xi^2, \: \: \:
\frac{\chi^{R_2}(T_{[g]})}{\dim R_2} \: = \: \xi.
\end{equation}
Again, we emphasize that the point of this section is merely to illustrate
the method in a simple well-known example.

\subsubsection{Twisted example:  ${\mathbb Z}_2 \times {\mathbb Z}_2$}

Let us apply the algorithm above to the case of
${\mathbb Z}_2 \times {\mathbb Z}_2$ with a twist,
as discussed in section~\ref{sect:twisted:z2z2:1}.

As discussed there, the center is one-dimensional, generated by
$\Pi$.

Now, suppose we are given the string amplitudes
\begin{equation}
Y_k \: = \: \frac{1}{|G|} \delta\left( \Pi \Pi^k T_{[1]} 
\right),
\end{equation}
and we want to compute the normalized characters of $T_{[1]}$.
(Clearly, this will be trivial, but $[1]$ is the only $\omega$-regular
conjugacy class, so for purposes of illustrating the method,
we will walk through this example.)
From~(\ref{eq:final-handle-id2}), we know that 
\begin{equation}
\frac{1}{|G|} \delta\left( \Pi^{k+1} T_{[g]} \right)
\: = \: \sum_R \left( \frac{ |G| }{\dim R} \right)^{2(k+1) - 2}
\left( \frac{ \chi^R(g) }{ \dim R} \right),
\end{equation}
which is a linear system of equations relating the normalized
characters $\chi^R(1)/\dim R$ to the $Y_k$ and the Vandermonde
matrix 
\begin{equation}
{\mathcal V}_{k,R} \: = \: \left( \frac{ |G| }{\dim R} \right)^{2k},
\end{equation}
and can be written in the form
\begin{equation}
\vec{Y} \: = \: {\mathcal V} \vec{\chi},
\end{equation}
where $\vec{\chi}$ is the vector of normalized characters
$\chi^R(1) / \dim R$ desired.

In the present case, ${\mathbb Z}_2 \times {\mathbb Z}_2$ with a twist,
there is only one irreducible projective representation, of dimension $2$,
hence
\begin{equation}
{\mathcal V}_{k,R} \: = \: \left( \frac{|G|}{2} \right)^{2k} \: = \:
2^{2k},
\end{equation}
so our system of equations is simply
\begin{equation}
Y_k \: = \: (2^{2k}) \left( \frac{\chi^R(1)}{\dim R} \right).
\end{equation}
(To be clear, this is many equations for one unknown, which is why
in general we restrict to a finite number of values of $k$.)

In principle this allows one to compute the normalized characters
in terms of the $Y_k$.  In this particular case, it is a fact that
$Y_k = 2^{2k}$, so we see that
\begin{equation}
\frac{\chi^R(1)}{\dim R} \: = \: 1,
\end{equation}
or simply,
\begin{equation}
\chi^R(1) \: = \: \dim R,
\end{equation}
a result which will not surprise the reader, but which will hopefully
help to illuminate the idea of the method.

\subsubsection{Twisted example:  $D_4$}

Now, let us apply these ideas to the case of $D_4$ with a twist,
using the computations in section~\ref{sect:twist:d4:1}.
Here, let us take the (two-dimensional)
center of the twisted gruop algebra to be
generated by $\{ T_{[b]} \}$, and use the 
string amplitudes (Dijkgraaf-Witten
correlation
functions) to compute the normalized characters and reproduce
the character table~\ref{table:td4-chars}.

As before, suppose we are given the string amplitudes
\begin{equation}
Y(\mu)_k \: = \: \frac{1}{|G|} \delta\left( \Pi T_{[b]}^k 
T_{\mu} \right),
\end{equation}
which are related to the normalized characters of $T_{\mu}$ by
\begin{equation}
Y(\mu)_k \: = \: \sum_R {\mathcal V}_{k,R} \frac{ \chi^R(T_{\mu}) }{ \dim R },
\end{equation}
for 
\begin{equation}
{\mathcal V}_{k,R} \: = \: \left( \frac{ \chi^R(T_{[b]}) }{ \dim R}
\right)^k.
\end{equation}
As there are only two irreducible projective representations, it suffices
to take $k \in \{0, 1 \}$ and write ${\mathcal V}_{k,R}$ as the
entries of a matrix
\begin{equation}
{\mathcal V} \: = \: \left[ \begin{array}{cc}
1 & 1 \\
+ (1+i)/2 & - (1+i)/2 \end{array} \right]
\: = \:
\left[ \begin{array}{cc}
1 & 1 \\
+\exp(i \pi/4)/\sqrt{2}  & - \exp(i \pi/4)/\sqrt{2} \end{array} \right].
\end{equation}

Using 
\begin{equation}
{\mathcal V}^{-1} \: = \: \frac{1}{2} \left[ \begin{array}{cc}
1 & +\sqrt{2} \exp(-i \pi/4) \\
1 & - \sqrt{2} \exp(-i \pi/4) \end{array} \right],
\end{equation}
one can then compute normalized characters from string amplitudes, 
formally as
\begin{equation}
\left( \frac{ \chi^R(T_{\mu}) }{ \dim R} \right) \: = \:
{\mathcal V}^{-1} \vec{Y}(\mu).
\end{equation}
For example, for $\mu = [b^3]$, the string amplitudes are
\begin{equation}
\vec{Y}([b^3]) \: = \: \left[ \begin{array}{c}
0 \\ 2 \end{array} \right],
\end{equation}
which implies
\begin{equation}
\left( \frac{ \chi^R(T_{[b^3]} }{ \dim R } \right)
\: = \: 
{\mathcal V}^{-1} \vec{Y} \: = \:
\left[ \begin{array}{c}
+(1-i) \\ -(1-i) \end{array} \right],
\end{equation}
correctly matching table~\ref{table:td4-chars}.

\subsubsection{Twisted example:  $S_n$}

In this section we discuss the symmetric group $S_n$ with discrete torsion.

First, let us describe the discrete torsion.
We can do this implicitly using the extension
$\tilde{S}_n$ presented in section~\ref{sect:ex:untwisted-tildes},
and comparing to the presentation of $S_n$ itself in
section~\ref{sect:ex:sn:1}.
Specifically, the extension is determined by an element of
$H^2(S_n,{\mathbb Z}_2)$, which maps into $H^2(S_n,U(1))$ and so
determines an element of discrete torsion.

We can compute the cocycle as follows, following \cite[pp. 9-10]{hh}.
Let 
$\theta : \tilde S_n \rightarrow S_n$ be the projection,
with kernel $\{ 1, z \}$, and let
$r$ be a section, meaning a map  $r: S_n \rightarrow \tilde S_n$, such that 
$\theta ( r ( a ) )  = a $ and $r (1) = 1$. 
A cocycle is given explicitly by 
\bea 
\alpha_r ( a , b ) = (-1)^{ n_r ( a, b ) }  ,
\eea
where 
\bea 
r(a) r(b) = z^{ n_r ( a , b ) }  r ( ab ) .
\eea
We can pick 
\bea 
&& \theta ( t_i ) = x_i, \cr 
&& \theta ( z ) = 1, \cr 
&& r ( x_i ) = t_i ~~~~~ r ( x_{ i_1} x_{ i_2}  \cdots  ) = t_{ i_1} t_{ i_2} \cdots 
\eea
The section can also be used to construct projective representations of $S_n$  
from the ordinary representations of $\tilde S_n$. 
Given representation matrices $R ( \tilde g )$  for $\tilde g \in \tilde S_n$, 
one gets projective representation matrices $P ( g )$ as
\bea 
 P ( g ) = R ( r (g) )
 \eea
as in \cite[Theorem 1.4]{hh}. 

As an example to illustrate the use of the above equations, consider the symmetric group $S_4$. It has three  generators $\{ x_1 , x_2 , x_3  \}$, which are the adjacent  transpositions $x_1 = (1,2) , x_2 = (2,3) , x_3 = (3,4)$.  The section $r $ is defined  by mapping words in the $x_i$ to words in $t_i$. 
As an example of cocycle factors deduced from the above equations, note that 
\bea
r( x_1 ) . r ( x_1 ) = t_1 t_1 = z = z r ( x_1^2 )  
\eea
Hence 
\bea 
\alpha ( x_1 , x_1 ) = (-1) .
\eea

Using the projection $\theta $ and the section $r$, the above equations specify a map from $ \mC ( \tilde{S}_4 ) $ to $ \mC_{ \omega } ( S_4 ) $. Using the character table for $ \tilde{S}_4$ (Table \ref{table:ts4-chars}), 
we note that the characters for elements in $\tilde{S}_4$, in the last three rows associated with non-trivial twist, and  corresponding to cycle structures $(2,1^2), (3,1)$ are zero. This means that the only non-zero $\omega$-regular classes in $ \mC_{ \omega } ( S_4 ) $ correspond to cycle structures  $ (1^4), (3,1) , (4)$. The equality of the number of $\omega$-regular conjugacy classes and irreducible projective reps illustrates our discussion of the center $\cZ ( \mC_{ \omega } ( G ) ) $: we observed that there is a basis for the center in terms of twist operators labelled by $\omega$-regular conjugacy classes and another basis in terms of projectors, labelled by irreducible projective representations. 
The splitting of $ ( 3,1) $ and $(4)$ into two columns illustrates the fact that characters are not class functions in the case of projective representations. Focusing on the column $(4)'$, and taking into account the dimensions of 
irreducible projective reps (given in the last three entries in the first column labelled by $(1^4)$), we find that the normalized characters $ \{ 1/\sqrt{2}  , -1/\sqrt{2}  , 0 \}$ distinguish the three irreducible projective reps.  
Following our discussion in section \ref{subsecPismall}, this means that a central element labelled by conjugacy class $(4)$ can be used to multiplicatively generate the center of $\cZ ( \mC_{ \omega } ( S_4 ) ) $.

\subsection{Algorithm for minimal generating subsets} \label{sec:fin-min-gen}

In the above, we have assumed we are given central elements which distinguish irreducible representations, or equivalently, 
multiplicatively generate the center $\cZ ( \mC_{\omega} ( G ))$. 
In this section, we outline an algorithm finding a minimal generating
subset of the center of a twisted group algebra, using the topological
field theory amplitudes. We start with a central element $C_a$. We can determine its normalized characters using the Burnside algorithm \cite{Burnside,Dixon,Schneider}, equivalently as explained earlier, by considering genus one amplitudes with insertions of boundaries labelled by $C_a$. If the number of distinct eigenvalues, i.e. the number of distinct normalized characters ${\chi_R ( C_a) \over \dim R } $ is equal to the dimension of $\cZ ( \mC_{\omega} ( G ))$, then we know that $C_a$ generates the center. 
But suppose the number of distinct eigenvalues is smaller. Let us ask how to determine
whether adding another central element $C_b$ 
indeed generates the center. 
This can be done by considering the structure constants of the multiplication operator 
for $C_a, C_b$ in the basis of central elements labelled by conjugacy class operators $T_{ \mu} $
\begin{eqnarray} 
C_a T_{ \mu } & = & (  {\mathfrak C}_{ a} )_{ \mu }^{ \nu} T_{ \nu },
\\
C_b T_{ \mu } & = & ( {\mathfrak C}_{ b} )_{ \mu}^{ \nu} T_{ \nu}.
\end{eqnarray}
These structure constants can be obtained from $G$-CTST amplitudes 
on the sphere:
\begin{eqnarray} 
 { 1 \over |T_{ \nu}| } \delta ( C_a T_{ \mu} T_{ \nu} )
& = & {\mathfrak C}_{ a \mu }^{ \nu },
\\
{  1 \over |T_{ \nu}| } \delta ( C_b T_{ \mu} T_{ \nu} )
& = &
 {\mathfrak C}_{ b  \mu }^{ \nu }.
\end{eqnarray} 
We know from~(\ref{eq:c-from-p}) that the projectors $P_R$ obey 
\begin{eqnarray} 
C_a P_R & = & { \chi^R ( C_a ) \over \dim R } P_R,
\\  
C_b P_R & = & { \chi^R ( C_b ) \over \dim R } P_R.
\end{eqnarray}
If $C_a, C_b$ generate the center, then the simultaneous eigenspaces of the matrices  
$( {\mathfrak C}_a ) , ( {\mathfrak C}_b )$   
are one-dimensional with eigenvalues 
\bea 
\left( { \chi^R ( C_a ) \over \dim R } ,   { \chi^R ( C_b ) \over \dim R }                    \right) . 
\eea

Motivated by AdS/CFT applications of minimal generating subspaces, 
we can start with the twist field associated to
(a representative of)
the smallest conjugacy class $T_{ a_1}$ 
(excluding the conjugacy class of the identity) 
and the associated structure constant matrix ${\mathfrak C}_{ a_1}$ 
obtained from $G$-CTST amplitudes involving $T_{ a_1}$, 
then alongside consider ${\mathfrak C}_{ a_2}$ 
for the next smallest conjugacy class. 
If the simultaneous eigenspaces are one-dimensional, 
we have a generating subspace spanned by $(T_{ a_1} , T_{ a_2}  )$. 
If the simultaneous eigenspaces are more than one-dimensional, 
we add another central element $T_{ a_3}$ and simultaneously diagonalize
${\mathfrak C}_{ a_1} , {\mathfrak C}_{ a_2} , {\mathfrak C}_{ a_3} $. 
If the eigenspaces are one-dimensional, then the ordered lists of eigenvalues of $ \{ {\mathfrak C}_{ a_1} , {\mathfrak C}_{ a_2} , {\mathfrak C}_{ a_3}  \}$ which give  
\bea 
\left\{ { \chi^R ( T_{ a_1}  ) \over \dim R } ,   
{ \chi^R ( T_{ a_2} ) \over \dim R }   , 
{ \chi^R ( T_{ a_3}  ) \over \dim R }   \right\} , 
\eea
can be used to label the irreducible representations. 

To find the eigenvalues for these basis elements in a 
minimal generating subspace, we have to solve the eigenvalue equations for 
the $K \times K $ matrices, where $K$ is the dimension of the center. 
For the characters of the remaining conjugacy classes, 
we use the inversion of Vandermonde matrices of smaller size as explained above.

\subsection{$G$-CTST and properties of characters of finite groups  }\label{sec:propchar} 

In this section we will use the properties of  handle-creation operators in 
$G$-CTST from section~\ref{sec:FTVac}, and the AdS/CFT-inspired construction 
of characters using minimal generating subspaces 
from the previous subsections   
\ref{subsecPismall}, \ref{sec:char-gen}, \ref{sec:fin-min-gen}, 
to derive certain integrality properties of residues of poles of
partition functions appearing in $G$-CTST.

Along the road to those physics results, we will derive some
mathematical properties of characters of finite groups.
We expect that these properties are already known in the mathematical literature; we are not claiming any fundamental mathematical novelty. 
We include them because they follow from the framework of $G$-CTST and are related to the properties of 
singularities in generating functions arising therein.  
The methods in the proof are based on the combinatorics of group multiplications along with linear algebra. 
Similar methods have been used to obtain integrality properties of characters 
in, for example, \cite{Chillag}. A comprehensive textbook discussion of these properties is in Chapter 3 of \cite{Isaacs}. 

In section \ref{sec:propcharI} we begin by deriving integrality properties for sums of 
characters of  a given conjugacy class 
$\cC_{ \mu}$, where the characters are being summed over certain restricted classes of irreducible representations.
The restrictions depend on the dimension of the irreducible representation or the character of certain additional specified conjugacy classes, where these conjugacy classes have the property that all their characters are integers. 
In section \ref{sec:propcharII} we extend the discussion to obtain  integrality properties of sums of powers of characters, where the sums are constrained by similar restrictions as in \ref{sec:propcharI}. We show that the integrality of these power sums is equivalent to factorisation properties of  polynomials arising in the Burnside algorithm  \cite{Burnside,Dixon,Schneider} for the computation of characters, 
which we will refer to as Burnside character polynomials. In section \ref{integGen} we show that the integer sums of normalized characters considered in \ref{sec:propcharI} and \ref{sec:propcharII} arise as residues of singularities in generating functions of $G$-CTST. 

For simplicity  we will restrict to Dijkgraaf-Witten
theories without discrete torsion (twisting) in this section.

\subsubsection{Integrality properties of some character sums }\label{sec:propcharI}

In this subsection we will derive some properties of characters that we
will use in the analysis of poles of $G$-CTST generating functions.

First, it is useful to rewrite \eqref{charbdyBurnside} with an  adjusted normalization
\bea\label{charbdyBurnside1}  
 {1  \over |G| } \delta ( \Pi  (|[g]| T_{[g]})^l  ) \: = \: 
\sum_{ R }\left (   {  |[g] | \chi^R ( g )  \over \dim R  }   \right )^l .
\eea
The ratios  $ |[g] | \chi^R ( g )  / \dim R $ in the right-hand side are known to be algebraic integers. 
This follows from the fact that eigenvalues of integer matrices (in this case, the matrix of structure constants of multiplication by the central elements $|[g]| T_{[g]}$ in  $ \cZ ( \mC ( G ) )$)   are algebraic integers 
(see e.g.~\cite[chapter 3]{Simon}).  
It is also known that algebraic integers form a ring. Hence a sum of algebraic integers is an algebraic integer.  
Thus, the sum 
\bea\label{normcharsum}  
\frac{ |[g]|  \chi^{\tilde  R'} (  g )
}{
\dim R' }
\: = \: \sum_{ R : R' }  {  |[g]|  \chi^R ( g )  \over \dim R  }    ,
\eea
(where the sum is over all the irreducible representations $R$ 
with a fixed $\dim R' = \dim R $ as in section~\eqref{sec:HandlesOnePt})
is an algebraic integer.  It is useful to rewrite  \eqref{OnePtChar} 
with the normalization
 \bea\label{OnePtChar1} 
  { 1 \over |G| }  \delta \left( \Pi^{ h }   |[g]| T_{ [g]}  \right)  \: =  \:
  \sum_{ R'} \left({  |G|^2 \over (\dim R')^2     } \right)^{h-1}
 { |[g]| \chi^{ \tilde R'} ( g ) \over \dim  R' } . 
  \eea
For the untwisted case $\mC ( G)$  the left-hand side gives a sequence of rational numbers for different values of $h$. 
In section \eqref{sec:HandlesOnePt}  we inverted the Vandermonde matrix of integers, applied it to a finite vector with the rational numbers on the left-hand side above, to give the characters   $ \chi^{ \tilde R'} ( g ) / \dim  R'$. 
Applying the same procedure here, we see that the normalized characters  
$ |[g]| \chi^{ \tilde R'} ( g ) / \dim  R' $
are rational numbers. 
Now, any algebraic integer which is rational is also integer 
(see e.g.~\cite[chapter III]{Simon}). 
This means that the sums of normalized characters in \eqref{normcharsum}  
are always integers, for any $\mC ( G ) $  (even though the individual terms in the sum may not be integers).  

To summarize, these arguments suggest the following
\vskip.1cm 
\noindent
{\bf Proposition 3.4.1-I:}  The sum of normalized characters 
\bea 
\sum_{ R : R' }  {  |[g]|  \chi^R ( g )  \over \dim R  } = { |[g]| \over \dim R' } \sum_{ R : R' } \chi^R ( g ) 
\eea
over all the irreducible representations $R$ of a fixed  dimension $\dim R'$  is an integer for any finite  group $G$. 

This is easy to verify 
in examples by inspection of finite group character tables. 
In addition, we expect that the statement above, 
as well as the other propositions in this section, 
likely already
exist
in the literature, though we are not able to give precise
references.  We include them here because we will use these results in the
analysis of poles of $G$-CTST generating functions.
We are not claiming any
fundamental mathematical novelty.

It is also known that the characters  $\chi^R ( g ) $ are algebraic integers 
(e.g.~\cite[chapter III]{Simon}), 
hence the sum 
\bea 
\chi^{ \tilde R' } ( g ) \equiv \sum_{ R : R' }  {   \chi^R ( g )  }
\eea
is an algebraic integer. 
The rationality of $ |[g]|  \chi^{\tilde R'}  ( g ) / \dim R' $ 
explained above also implies that 
${   \chi^{\tilde R'}  ( g )    }$ is rational. Using again the fact that rational algebraic integers are integers, we conclude that $\chi^{ \tilde R' } ( g )$ are integers. 
We state this as 
\vskip.1cm 
 \noindent
{\bf Proposition 3.4.1-II:} The sum of the characters 
\bea 
\sum_{ R :R' }   \chi^R ( g )   
\eea
over all irreducible representations $R$ with a fixed dimension $\dim R'$,
is an integer, for any finite group $G$.

 A corollary of the discussion on integrality of character sums above, is that 
 if for every irreducible representation $R$   of a finite group $G$ which has a unique value of the dimension, i.e.~a value not shared by any other irreducible representation, the characters 
${|[g]|  \chi^R (g) \over \dim R }$ and $ \chi^R ( g ) $ are integers for $g$ in any conjugacy class.

Following the discussion in  section \ref{sec:char-gen} where we consider linear systems for a given ${ \chi^R ( T_{\mu}  ) \over \dim R }  $ using  a pair of central elements, we can generalize the above argument. We start again with the untwisted case $\mC ( G )$. Consider central elements $ \{ C_1 , C_2 \} $, chosen to have the property that ${ \chi^R ( C_1 ) \over \dim R   }$ and 
${ \chi^R ( C_2 ) \over \dim R } $ are both integers for all $R$.  We do not require here that $ C_1, C_2 $ generate the center
$\cZ ( \mC ( G ))  $ in the present discussion.  The key equation  is \eqref{C1C2sys}, part of which we repeat for convenience,  is 
\bea\label{C1C2sys1} 
{ 1 \over |G| } \delta \left ( \Pi C_1^{ k }  ( C_2^l   |C_{ \mu} | T_{ \mu } )  \right ) 
 = \sum_{ R }\left( { \chi^R ( C_1 ) \over \dim R }  \right )^{ k  }  \left( { \chi^R ( C_2 ) \over \dim R } \right)^{ l } { \chi^R ( |C_{\mu} |  T_{\mu  } ) \over \dim R } .
\eea
It is worth noting that the product in the sum above, namely 
\begin{equation}
 \left( { \chi^R ( C_2 ) \over \dim R } \right)^{ l } { \chi^R ( |C_{\mu} |  T_{\mu  } ) \over \dim R } ,
\end{equation}
is a product of algebraic
integers and hence itself an algebraic integer.
Using the discussion in  section \ref{sec:char-gen}, we can construct the character sums 
\bea 
\sum_{ R : [ C_1  , C_2 ] } { \chi^R ( |C_{\mu} |  T_{\mu  } ) \over \dim R }
\eea 
where $ R $ is being summed over all the irreducible representations 
having a fixed pair of eigenvalues for $[C_1, C_2]$, 
using inverses of integer Vandermonde matrices multiplying the combinatoric data on the left-hand side of \eqref{C1C2sys1} consisting of rational numbers. 
Thus we conclude that these sums, which are known to be algebraic integers,  
are in fact integers. This also means that the character of a $\chi^R ( g ) $ of a group element $g \in C_{\mu} $ is rational, and since  it is known to be an algebraic integer, also  in fact integer. 
By taking $C_1$ to be  the handle creation  operator with eigenvalues 
${ |G|^2 \over (\dim R)^2 } $ 
and $C_2$ the sum of elements in  a conjugacy class $ \mathcal{ C}$  
with the property that $ \chi^R ( g ) $ for $g \in \mathcal{ C} $  
is an integer for all irreducible representations $R$, we conclude
\vskip.1cm 
\noindent 
{\bf Proposition 3.4.1-III:} The character sums  
\bea 
\sum_{ R :  [ \dim R'    , \chi^{R''}  ( \mathcal { C }  )  ] } { \chi^R ( |C_{\mu} |  T_{\mu  } ) \over \dim R }
\eea 
and 
\bea 
\sum_{ R :  [ \dim R'    , \chi^{R''} ( \mathcal { C }  )  ] } { \chi^R ( g  ) }  \hbox { for } g \in \mathcal { C }_{ \mu} 
\eea
for any conjugacy class $\mathcal{ C}_{ \mu}   $, 
over irreducible representations with a fixed specified dimension denoted 
$\dim R'$ and a fixed value of the character for the conjugacy class 
$\mathcal{ C } $, are integers.  

If we take $ [ C_1, C_2 ] $ to be two conjugacy classes having integer characters, then we have 
\vskip.1cm 
\noindent 
{\bf Proposition  3.4.1-IV:} 
The character sums 
\bea 
\sum_{ R :  [  \chi^{R_1} ( \mathcal { C }_1  )   , \chi^{R_2}  ( \mathcal { C }_2  )  ] } { \chi^R ( |C_{\mu} |  T_{\mu  } ) \over \dim R }
\eea 
and 
\bea 
\sum_{ R :  [ \chi^{R_1}  ( \mathcal { C }_1  )   , \chi^{R_2}  ( \mathcal { C }_2   )  ] } { \chi^R ( g  ) }  \hbox { for } g \in \mathcal { C }_{ \mu} 
\eea
for any conjugacy class $\mathcal { C }_{ \mu} $ are integers, where the sum is  over all  irreducible representations which have  fixed characters  $ [ \chi^{R_1} ( \mathcal { C }_1  )   , \chi^{R_2}  ( \mathcal { C }_2  )  ] $
 for two conjugacy classes $\mathcal { C }_1  , \mathcal { C }_2$, and  where these latter are conjugacy classes known to have  integer characters for all irreducible representations $R$. This property for $\mathcal { C }_{ \mu} $
 generalizes to the case where we fix the characters for any number of conjugacy classes $\{ \mathcal { C }_1  , \mathcal { C }_2 , \cdots , \mathcal {C}_m \} $ having the property that all their irreducible characters are integers. We also have this integrality property for $\mathcal { C }_{ \mu} $ when we fix 
 $\{ \dim R, \mathcal { C }_1  , \mathcal { C }_2 , \cdots , \mathcal {C}_m \} $. 

 Integrality properties of fusion matrices  and quantum dimensions have  recently been studied using Galois theory methods \cite{Matt2109} in the context of 3D topological  quantum field theories.  The combination of Galois theory methods with the constructive methods used here in general classes of topological field  theories would be an interesting area for future investigation. 

\subsubsection{Integrality of power sums and factorisation properties of character polynomials }\label{sec:propcharII} 

In the previous subsection, as part of our physical analysis of $G$-CTST,
we derived some intermediate mathematical integrality properties
involving single characters.  In this subsection we similarly derive integrality
properties 
for power sums of characters which have implications for the factorization
properties of the Burnside character polynomials.
In the next subsection we will
apply these properties 
to the analysis of generating 
functions in $G$-CTST.

For a conjugacy class $\cC_{ \mu} $ consider a diagonal matrix $X_{ \mu} $ of size $K$, with entries 
${  |\cC_{ \mu} |  \chi^R ( g ) \over \dim R }  $ for $ g \in \cC_{ \mu} $ where $K$ is the  number of conjugacy classes in $G$. The determinant $\det ( x - X_{ \mu}  )$ is a polynomial in $x$ 
\bea 
\det (x -  X_{\mu}  ) & = & x^K - x^{ K -1} \tr X +   \cdots + (-1)^K \det X_{ \mu},
\nonumber \\ 
 & = & \sum_{ i =1}^{ K } (-1)^i x^{ K - i } e_{ i } ( X ) 
\eea
where $e_{ i } ( X )$ are elementary symmetric polynomials. 
They can be expressed in terms of traces of $X$ and in terms of the eigenvalues of $x_i$ of $X$ as 
\bea\label{ekformula}  
 e_{ k  } ( X ) & = & \sum_{ p \vdash k  } { ( -1)^{ k    - \sum_{ i } p_i } \over \prod_{ i } i^{ p_i } p_i! }  \prod_i ( \tr X^i )^{ p_i },
\nonumber \\ 
& = & \sum_{ 1 \le i_1 < i_2 < \cdots < i_p \le n } x_{ i_1} x_{ i_2} \cdots x_{ i_k  } 
\eea
Here $p$ is a partition of $k$, with $p_i$ parts of length $i$, so that  $ \sum_{ i } i p_i = k $. 
As reviewed in \cite{deMelloKoch:2021lqp} the quantity $ \det ( x - X_{ \mu} ) $, viewed as a polynomial in $x$, is also the characteristic polynomial for the integer matrix ${  |\mathcal{C}_{ \mu}| | \mathcal{C}_{\nu}| \over |\mathcal{C}_{ \lambda } | }( \mathfrak{C}_{ \mu } )_{ \nu }^{ \lambda } $ of structure constants of $\cZ ( \mC ( G ))$.  Solving for the eigenvalues of the matrix of structure constants for conjugacy classes $\cC_{ \mu} $ is a step in determining the character table in the Burnside algorithm \cite{Burnside}. A useful piece of terminology is that $ \det ( x - X_{ \mu} ) $ is an integer monic polynomial: a monic polynomial has the coefficient of the highest power of $x$ to be equal to $1$ while all the other coefficients are also integers.

The above arguments  for integrality of sums of characters apply equally well for the power sums. In this case we consider, for fixed $k$ and for 
$h \in \{ 0 , 1 , \cdots , K -1 \}$ 
 \begin{eqnarray} 
\lefteqn{
  { 1 \over |G| }  \delta \left( \Pi^{ h }   ( |[g]| T_{ [g]} )^k   \right)
} \nonumber \\
 & = & 
  \sum_{ R} \left({  |G|^2 \over (\dim R)^2     } \right)^{h-1}
\left ( { |[g]| \chi^{ R} ( g ) \over \dim  R } \right )^k,
\nonumber \\   
& =  & \sum_{ R'} \left({  |G|^2 \over (\dim R')^2     } \right)^{h-1}
\sum_{ R : R'} \left ( { |[g]| \chi^{ R} ( g ) \over \dim  R } \right )^k .
\label{OnePtChar2}
  \end{eqnarray}
  The last line includes a sum over irreps $R$ having a fixed dimension $\dim R'$.  This allows us to write, in terms of an inverse Vandermonde matrix, the power sums over irreducible representations  $R$ of $G$ 
with a fixed dimension $\dim R = \dim R'$ 
 \bea 
 \sum_{ R : R' }   \left ( {  |[g]|  \chi^R ( g )  \over \dim R  } \right )^k .
 \eea
These are known, on general grounds, to be algebraic integers.
Applying the reasoning in section \ref{sec:propcharI} above 
to these power sums,
they can be expressed as a matrix product of a rational matrix (inverse of a Vandermonde matrix) times a vector of rational numbers (obtained from the evidently  rational numbers on the LHS of \eqref{OnePtChar2}). This means that these sums of powers, restricted to all  irreducible representations $R$ having the same dimension as $R'$, are actually integers.

It is now useful to consider a diagonal matrix $X_{ \mu}^{ (R')} $  of size equal to the number $K^{ (R')} $ of distinct irreducible representations with the same dimension as $R'$, and with entries equal to ${  |[g]|  \chi^R ( g )  \over \dim R  }$ as $R$ ranges over the distinct $R$ with the specified dimension. We can construct a polynomial $ \det ( x - X_{ \mu}^{ (R')}  )$ of degree $K^{(R')}$. The coefficients of the powers of $x$ are elementary symmetric polynomials $e_i ( X^{(R')}  ) $, expressible as polynomials in these  normalized characters ${  |[g]|  \chi^R ( g )  \over \dim R  }$ for $R$ having fixed dimension $\dim R'$. Since these normalized characters are known to be algebraic integers, the elementary symmetric polynomial functions of these (which are sums of products of these according to the second line in \eqref{ekformula}) are algebraic integers. These elementary symmetric polynomials are also expressible in  terms of linear combinations with rational coefficients of power sums (first line of  \eqref{ekformula}). These power sums are integers as explained above. Combining these facts, and since numbers which are rational and algebraic integer are also integers, we conclude that these coefficients of powers of $x$ in $ \det ( x - X_{ \mu}^{ (R')}  )$
are actually integers. Thus $ \det ( x - X_{ \mu}^{ (R')}  ) $ is an integer monic polynomial in the variable $x$.  
Since the diagonal entries of the diagonal matrix $X_{ \mu}^{ (R')} $ form a subset of the entries of the diagonal matrix $X_{ \mu} $ defined above,  we see that $ \det ( x - X_{ \mu}^{ (R')}  )$ is an integer monic polynomial which is a factor of the Burnside character polynomial $ \det ( x - X_{ \mu } ) $. We summarise this conclusion as 
  \vskip.2cm 
\noindent 
{\bf Proposition 3.4.2-I:  }  
 The Burnside character polynomial for any conjugacy class $\cC_{ \mu} $, which is an integer monic polynomial,  factorises into lower degree integer monic polynomials parametrised by the  list of distinct dimensions $\dim R'$ 
\bea 
\det ( x - X_{ \mu}  ) = \prod_{ R'} \det  ( x - X_{ \mu}^{ (R')}  ) .
\eea

Following the discussion in section 3.3.1, we can also consider further integrality properties for powers of normalised characters summed over sets of irreps restricted by dimension as well as characters of conjugacy classes.
By following the argument above, this integrality of power sums leads to more refined factorisation properties of the Burnside character polynomials.  Suppose $\cC_1 $ is a conjugacy class with integer characters. i.e. for all irreducible representations $R$ of $G$, the characters $\chi^R ( g ) $ for $  g \in \cC_1$ are integers. Let $\chi^{ \cC_1 ; R_1' }$  be the list of the distinct values of these characters, and $ K^{ \cC_1 ; R_1' } $ be the multiplicity of the eigenvalue. We have 
$\sum_{ R_1'} K_{ \cC_1 ; R_1'} = K$. Let $X_{ \mu}^{ ( \cC_1 ; R_1')} $ be the diagonal matrix with entries 
${ \chi^R ( \cC_{ \mu} ) \over \dim R } $  for irreducible representations $R$ having 
\bea 
 \chi^R ( g ) = \chi^{ R_1'} ( g )   \hbox { for } g \in \cC_1 ,
\: \: \:
 \chi^R ( g  ) = \chi^{ \cC_1 ; R_1' } .
\eea
The polynomial $\det ( x - X_{ \mu}^{ \cC_1 ; R_1' } ) $ is an integer monic polynomial. 
\vskip.2cm 
\noindent 
{\bf Proposition 3.4.2-II:  }  
 The Burnside character polynomial for any conjugacy class $\cC_{ \mu} $, which is an integer monic polynomial,  factorises into lower degree integer monic polynomials parametrised by the  list of distinct characters 
 $\chi^{ \cC_1 ; R_1' } $
\bea 
\det ( x - X_{ \mu}  ) = \prod_{  R_1'} \det ( x - X_{ \mu}^{ \cC_1 ; R_1' } ) .
\eea

Let the pair $ [R' , R_1' ]  $  be labels for pairs of irreducible representations which run over the distinct possible values of 
$ [ \dim R , \chi^R ( g ) ] $ for $g \in \cC_1$. Let $ K^{ \Pi , \cC_1 ; R' , R_1' } $ be the multiplicity of the pair of values associated with $ [R' , R_1' ]  $. For any other conjugacy class $ \cC_{ \mu} \ne \cC_1$, we can construct the integer monic polynomial $\det ( x - X_{\mu}^{ \Pi , \cC_1 ; R' , R_1' } ) $  of degree $K^{ \Pi , \cC_1 ; R' , R_1' } $. 
We have the factorisation property 

\vskip.2cm 
\noindent 
{\bf Proposition 3.4.2-III:  }  
The Burnside character polynomial for any conjugacy class $\cC_{ \mu} $, which is an integer monic polynomial,  factorises into lower degree integer monic polynomials parametrised by the  list of distinct ordered pairs 
 $[ \dim R' , \chi^R ( g ) ] $ for $ g \in \cC_1 $ 
\bea 
\det ( x - X_{ \mu}  ) = \prod_{ R ,  R_1'} \det ( x - X_{ \mu}^{ \Pi , \cC_1 ; R' ,  R_1' } ).
\eea

These factorisation properties can further be generalised to run over lists \linebreak 
$[ \dim R'  , \chi^R ( g_1 ) , \cdots , \chi^R ( g_m ) ] $ for $ g_1 \in \cC_1 , g_2 \in \cC_2 , \cdots , g_{ m } \in \cC_m $ 
where $\cC_1 , \cC_2 , \cdots , \cC_m $ have integer characters. We can also drop $\dim R'$ from the lists to have factorisation over distinct lists $ [ \chi^R ( g_1 ) , \cdots , \chi^R ( g_m ) ]$.

\subsubsection{Integral power sums as residues of singularities in $G$-CTST generating functions }
\label{integGen} 

In this subsection we now apply the properties we have derived to the
analysis of $G$-CTST generating functions.

We observe that the integer sums of normalized characters and sums of powers of normalized characters  derived in sections \ref{sec:propcharI} and \ref{sec:propcharII} arise as residues at singularities of $G$-CTST generating functions. The argument is an extension of the one in section 5 of \cite{deMelloKoch:2021lqp}. 
Let us define a sum over arbitrary numbers of handles of the string amplitude with one boundary labelled by conjugacy class $\cC_{ \mu} $  \eqref{OnePtChar1} weighted by the appropriate power of the string coupling. 
Taking $ g \in \cC_{ \mu} $, i.e. $ [g] = \cC_{ \mu} $ we write  
\begin{eqnarray} 
\lefteqn{
 g_{ st}^{-1}  Z ( g_{st} ; \cC_{ \mu} )
} \nonumber \\
&  = & \sum_{ h =0 }^{ \infty } {g_{ st}^{ 2h -2 } \over |G| }  \delta \left( \Pi^{ h }   |[g]| T_{ [g]}  \right)  \: =  \:
  \sum_{ h } \sum_{ R'} \left({  g_{st}^2  |G|^2 \over (\dim R')^2     } \right)^{h-1}
 { |[g]| \chi^{ \tilde R'} ( g ) \over \dim  R' },
\\ 
 & = & \sum_{ R'} { 1 \over ( 1 - g_{ st}^2 |G|^2 /(\dim R')^2 ) }  { |[g]| \chi^{ \tilde R'} ( g ) \over \dim  R' }.
\end{eqnarray}
The poles of this generating function are at 
\bea 
g_{ st}= {  ( \dim R') \over |G|} ,
\eea
and the residues are 
\bea 
{ |[g]| \chi^{ \tilde R'} ( g ) \over \dim  R' } = { |[g]| \over \dim R' } \sum_{ R : R' } \chi^R ( g ) ,
\eea
which we showed to be integers (proposition 3.4.1-I). 
Similarly we can define a stringy generating function 
for the $k$'th  power sums 
\begin{eqnarray} 
\lefteqn{  g_{ st}^{k-2}  Z ( g_{st} ; \cC_{ \mu} , k  )
} \nonumber \\
& =  &
 \sum_{ h =0}^{ \infty  }  { g_{st}^{ 2h-2}  \over |G| }  \delta \left( \Pi^{ h }   ( |[g]| T_{ [g]} )^k   \right),
\\
   & = & 
 \sum_{ h=0}^{ \infty}  \sum_{ R} \left({  g_{st}^2 |G|^2 \over (\dim R)^2     } \right)^{h-1}
\left ( { |[g]| \chi^{ R} ( g ) \over \dim  R } \right )^k,
\\  
& = & \sum_{ h =0}^{ \infty }  \sum_{ R'} \left({  g_{st}^2 |G|^2 \over (\dim R')^2     } \right)^{h-1}
\sum_{ R : R'} \left ( { |[g]| \chi^{ R} ( g ) \over \dim  R } \right )^k,
\\  
& = & \sum_{ R'} { 1 \over ( 1 - g_{st}^2 |G|^2 /( \dim R')^2 )  } \sum_{ R : R'} \left ( { |[g]| \chi^{ R} ( g ) \over \dim  R } \right )^k .
\end{eqnarray}
The singularities are at 
\bea 
g_{st} = {  ( \dim R' ) \over |G| } ,
\eea
while the respective residues are 
\bea 
\sum_{ R : R'} \left ( { |[g]| \chi^{ R} ( g ) \over \dim  R } \right )^k .
\eea
As shown in proposition 3.4.1-II these residues of the $G$-CTST generating function defined are integers. 

The connection between integer  character sums and residues of $G$-CTST partition functions  extends to the more refined sums considered in sections \ref{sec:propcharI} and \ref{sec:propcharII}. As an example consider \eqref{C1C2sys1} involving powers of two conjugacy class sums $C_1 , C_2$ and a single power of $\cC_{ \mu} $  and let us introduce a partition function depending on two chemical potentials $\mu_1 , \mu_2$ 
\begin{eqnarray}   
\lefteqn{Z ( \mu_1  , \mu_2 ; C_1 , C_2 ) }
\nonumber \\
&  =  &
\sum_{ k , l =0}^{ \infty } \mu_1^{ k }  \mu_2^{ l }   { 1 \over |G| } \delta \left ( \Pi C_1^{ k }  ( C_2^l   |C_{ \mu} | T_{ \mu } )  \right ),
\nonumber \\
&  = & \sum_{ k , l =0}^{ \infty } \sum_{ R }\left(  { \mu_1 \chi^R ( C_1 ) \over \dim R }  \right )^{ k  }  \left( { \mu_2 \chi^R ( C_2 ) \over \dim R } \right)^{ l } { \chi^R ( |C_{\mu} |  T_{\mu  } ) \over \dim R },
\nonumber \\  
& = & \sum_{ R } { 1 \over ( 1 - { \mu_1 \chi^R ( C_1 ) \over \dim R } )}
 { 1 \over ( 1 - { \mu_2 \chi^R ( C_2 ) \over \dim R } )} { \chi^R ( |C_{\mu} |  T_{\mu  } ) \over \dim R },
\nonumber \\
 & = & \sum_{ R_1 , R_2 }  { 1 \over ( 1 - { \mu_1 \chi^{R_1}  ( C_1 ) \over \dim R_1 } )}
 { 1 \over ( 1 - { \mu_2 \chi^{R_2}  ( C_2 ) \over \dim R_2 } )} 
 \sum_{ R : [ \chi^{R_1} ( C_1 ) , \chi^{ R_2} ( C_2 ) ] }  { \chi^R ( |C_{\mu} |  T_{\mu  } ) \over \dim R } .
 \label{ZC1C2sys1} 
\end{eqnarray}
In the last line, we have introduced sums over a complete set of pairs  of irreducible representations $R_1 , R_2 $ which have distinct character values $[ \chi^{R_1} ( C_1 ) , \chi^{ R_2} ( C_2 ) ]$. For each pair of values, we have a sum over $R$ running over the distinct irreducible representations having these characters. It follows that the singularities of these generating functions are at 
\begin{equation} 
 \mu_1 = { \dim R_1 \over \chi^{ R_1} ( C_1 ) },
\: \: \
 \mu_2 = { \dim R_2 \over \chi^{ R_2} ( C_2 ) } . 
\end{equation}
The residues at these simgularities are 
\bea 
\sum_{ R : [ \chi^{R_1} ( C_1 ) , \chi^{ R_2} ( C_2 ) ] }  { \chi^R ( |C_{\mu} |  T_{\mu  } ) \over \dim R } .
\eea
These residues are integers as explained in Proposition 3.3.2-IV.

\section{Further remarks on  $G$-CTST and future directions }\label{sec:GCTST}

We collect a few comments here on the stringy interpretation of the determinants that have played a central role in the algorithms earlier in the paper. We find a link to plethystic exponentials of low genus amplitudes. The plethystic exponential function has well known applications in AdS/CFT relating the counting of single trace gauge invariants in CFT to multi-trace counting \cite{HeHananyPexp} . It also has a related application in tensor model holography, relating the counting of connected and disconnected surfaces which are related to tensor model invariants \cite{JBSR}. Careful quantum gravitational discussions of the normalizations of partition functions relevant to combinatoric topological strings are in \cite{MarMax,Gardiner:2020vjp,Moore}. 

The second point we develop is $S$-duality for $G$-CTST. While $S$-duality was discussed in \cite{deMelloKoch:2021lqp} in terms of entangled disconnected surfaces, we observe that there is also an interpretation of the $S$-dual amplitudes in terms of the inversion of the handle-creation operator in the group algebra of $G$. We observe that for both the untwisted and untwisted case this inverse operator is well-defined.  We give an expression for the inverse handle creation operator as an expansion in the projector basis $\cZ ( \mC_{ \omega}  ( G ) ) $. A combinatoric description of the expansion  in terms of the conjugacy class basis for $\cZ ( \mC_{ \omega}  ( G ) ) $ is an interesting question. The third point concerns the implications of finiteness of $G$ for relations between $G$-CTST amplitudes.

\subsection{Construction of integer ratios $|G|^2/(\dim R)^2$ and stringy interpretation}\label{sec:DetDiscon} 

\subsubsection{Background}

The construction of the integer ratios $ |G|/\dim R $ from group multiplications in \cite{deMelloKoch:2021lqp} used  the determinant $\det  ( x - X   )$, and its expansion in terms of products of traces
\bea\label{detAlgo} 
\det  ( x - X   ) =  x^K  - e_1 ( X ) x^{ K -1} + e_2 ( X ) x^{ K -2} + \cdots + (-1)^K e_K ( X ),
\eea
where the $e_i$ denote the elementary symmetric functions, given by
\bea
&& e_0 ( X ) = 1, \cr
&& e_1 ( X  ) = \sum_{ i } X_{ i },   \cr
&& e_2( X ) = \sum_{ 1 \le i < j \le  K } X_{ i } X_{ j } , \cr
&& e_{ l } ( X ) = \sum_{ 1 \le i_1 <  i_2 <  \cdots <  i_l  \le K } X_{ i_1} X_{ i_2} \cdots X_{ i_l }.
\eea
The elementary symmetric functions can be expressed in terms of traces of $X$ as in \eqref{ekformula}. 

As was argued in \cite{deMelloKoch:2021lqp}, the algorithm presented there
was a stringy construction more than a field theoretic construction, 
since it involved combining amplitudes of different genera, but there was not
a crisp simple connection between the algorithm and a stringy observable.

As a first step in this direction, note that $e_1(X)$ is $ \tr X = Z_{ h = 2}$. In a stringy partition function this  is naturally weighted with
$g_{st}^2$. The next elementary symmetric polynomial,
$e_2(X)$, is a linear combination of $ \tr X^2 = Z_{ h =3}  $ and 
$ (\tr X)^2 = Z_{ h=2}^2$. 
Both of these are weighted with $g_{st}^4$. 
The next elementary symmetric polynomial,
$e_3 (X)$, is a linear combination of $Z_{ h=4}, Z_{ h=3} Z_{ h=2}$, and
$Z_{ h=2}^3$, all of which are naturally weighted by $g_{st}^6$. 
In general, $e_k ( X ) $ is associated with $g_{st}^{2k}$.

The determinant above can be written as
\bea 
\det ( X - x ) = x^K \left ( 1  - e_1 ( X ) x^{ -1} + e_2 ( X ) x^{  -2} + \cdots + (-1)^K  x^{ - K } e_K ( X ) 
\right ) .
\eea
By substituting $ x \rightarrow g_{ st}^{-2} $, we can write
\bea 
x^{ - K } \det ( X - x ) \vert_{ x = g_{st}^{ -2} } 
= 1 - e_1 ( X ) g_{st}^2 + e_2( X ) g_{ st}^4 + \cdots + (-1)^k g_{ st}^{ 2K } e_K ( X ).
\eea
This looks like a stringy observable. We develop a link with disconnected string diagrams below. 

\subsubsection{Determinant from generating function of disconnected worldsheets   }

Start with the observation that a generating function of  disconnected diagrams of genus $2$ or higher can be obtained by expanding the exponential of a sum
\begin{eqnarray}
\cZ_{ \rm{ disconn} } ( g_{st}^2  ) & = & 
 \exp { \sum_{ k=1}^{ \infty } g_{st}^{2k} { \cZ_{ k+1}  \over k }  }   =
\exp { \sum_{ k=1}^{ \infty } g_{st}^{2k} { \tr X^{ k} \over k }  },
\\
& = & \prod_{ k =1 }^{ \infty } \sum_{ p_k =0 }^{ \infty } { g_{ st}^{ 2k p_k }  \over p_k! }  \left ( { \tr X^{ k} \over k } \right  )^{ p_k },
\\
& = & \sum_{ p_k =0 }^{ \infty }   \prod_{ k =1 }^{ \infty }  { g_{ st}^{ \sum_k 2k p_k } \over k^{ p_k} p_k! }
 ( { \tr X^{ k}} )^{ p_k }.
\end{eqnarray}
The argument of the exponential is motivated by the plethystic exponential function as studied in \cite{HeHananyPexp,JBSR}. Now observe that the first line is a determinant:
\bea  \label{eq:pe}
\exp { \sum_{ k=1}^{ \infty } g_{st}^{2k} { \tr X^{ k} \over k }  }
\: = \:
 \exp {  \left (  -  \tr \log ( 1 - g_{st}^2 X ) \right ) } 
\: = \:
 { 1 \over  \det ( 1 - g_{st}^2 X ) },
\eea
where in the last equality, we used 
\bea
\det ( A ) \: = \: \exp { \tr \log ( A ) }.
\eea
We conclude
\bea\label{detDisconn}
\det ( 1 - g_{st}^2 X ) \: = \: { 1 \over \cZ_{ \rm{ disconn} } ( g_{st}^2  ) }.
\eea

So the determinant used in the algorithm for $|G|^2/(\dim R)^2$ is nothing 
but the inverse of the generating function for the disconnected diagrams. 
The zeroes of this inverse generating function are at 
$g_{st}^2 = (\dim R)^2/|G|^2 $, or
$g_{st}^{-2}  = |G|^2/(\dim R)^2$. 
A remarkable fact is that this inverse generating function truncates at a finite power of $g_{st}^2 $. 
This is due to the finiteness properties of the theory.  Another way to express the remarkable fact is that the generating function of disconnected string diagrams is a rational function.

In \cite{deMelloKoch:2021lqp}, it was observed that finding the zeroes of $\det  ( x - X   )$ in \eqref{detAlgo},  
viewed as a function of $x$, gives a finite algorithm (which uses as input the products of traces of $X$ available from $G$-CTST partition functions) to arrive at the integer ratios $ |G|/\dim R$. The identification of the formal variable $x$ with $g_{st}^{-2}$ above and the equation  \eqref{detDisconn}  shows that the integer ratios have the physical interpretation of being the locations in the $ g_{st}^{-2} $ plane of the poles of the generating function 
of disconnected amplitudes. It was also observed in \cite{deMelloKoch:2021lqp} that the poles of the connected generating function as a function of $g_{st}$  are given in terms of the integer ratios $|G|/\dim R$.  Connected and disconnected generating functions are related through the plethystic exponential function (see \cite{HeHananyPexp} for applications of the plethystic exponential in the combinatorics of moduli spaces of supersymmetric gauge theories).

\subsection{$S$-duality  in $G$-CTST and the inverse handle-creation operator   }

Following the discussion in \cite{deMelloKoch:2021lqp},  the generating function of connected closed string amplitudes is 
\bea 
Z ( g_{ st} ) && =  \sum_{ h =0}^{ \infty } g_{ st}^{ 2h -2 } \sum_{ R } { ( |G| )^{ 2h -2 } \over (\dim R )^{ 2h -2 } },
\\  
&& = \sum_{ R } { (\dim R)^4 \over g_{st}^2 |G|^2 ( (\dim R)^2 - |G|^2 g_{st}^2 ) } . 
\eea
An $S$-dual generating function is defined as 
\bea 
&& \tilde Z ( \tilde g_{st} ) = - \tilde g_{ st}^{ -4} Z ( g_{st} \rightarrow \tilde g_{ st}^{ -1} ) .
\eea
It is calculated to be 
\bea\label{Sdualexp}  
\tilde Z ( \tilde g_{st} ) &&= \sum_{ R }  { ( \dim R )^4 \over |G|^4 }  \left( 1 - \tilde g_{ st}^2{  (\dim R)^2 \over |G|^2 }\right)^{ -1},   
\\
&& = \sum_{ R }  { ( \dim R )^4 \over |G|^4 } \left( 1 + \tilde g_{ st}^2 { ( \dim R)^2 \over |G|^2 } + \tilde g_{ st}^4 {  (\dim R)^4 \over |G|^4 } + \cdots \right),  
\\
&& = \sum_R \sum_{ k =1}^{ \infty } \tilde g_{ st}^{ 2 + 2 k }  {  (\dim R)^{2+2k}    \over |G|^{ 2 + 2k } }. 
\eea
In \cite{deMelloKoch:2021lqp} a geometrical interpretation for the positive power sums of dimensions was given in terms of disconnected entangled surfaces. Here we develop an alternative interpretation of this $S$-dual expansion. 

Recall the handle creation operator 
\bea 
\Pi = \sum_R { |G|^2 \over (\dim R )^2 } P_{ R } 
\eea
with the property $ \delta ( P_R ) =  { (\dim R)^2\over |G| }  $ so that the genus $h$ partition function 
is obtained by taking the trace of $h$ powers of $\Pi$. 
\bea 
Z_h = { 1\over |G| } \delta ( \Pi^h ) = \sum_{ R } { ( \dim R )^{ 2h -2 } \over |G|^{ 2h -2 }} .
\eea
We observe that the handle creation operator has an inverse element in the center of the group algebra, which is given by  
\bea
\Pi^{-1} = \sum_{ R } { (\dim R)^2 \over |G|^2 } P_R .
\eea
We have 
\bea 
\Pi \Pi^{ -1} && = \sum_{ R , S } { (\dim R)^2 \over |G|^2 }{ |G|^2 \over (\dim S)^2 } P_{ R } P_{ S },
\\ 
&& = \sum_{ R } P_{ R } = 1 .
\eea
We propose to interpret the inverse handle creation operator $\Pi^{-1}$ as the handle creation operator of the 
$S$-dual theory and denote it as $ \Pi^{-1} =  \tilde \Pi$. 

Note that the leading order term in the $S$-dual generating function  \eqref{Sdualexp} is
\bea 
{ 1 \over |G| } \delta ( \tilde \Pi ) &&  = \sum_{ R } { 1\over |G| } { (\dim R)^2 \over |G|^2 } \delta ( P_R ),
\\  
&& = \sum_{ R } { (\dim R)^4 \over |G|^4 } .
\eea
Since there is a single power of $\tilde \Pi$, it is natural to interpret this as the partition function at genus one of the $S$-dual theory. The higher powers are 
\bea 
{ 1 \over |G| } \delta ( \tilde \Pi^{ k }  ) = \sum_{ R }  { (\dim R)^{ 2 + 2k }  \over |G|^{2+2k}  } 
\eea
which can therefore be interpreted as genus $k$ partition function of the dual theory.

{\bf Remark:}  It would be interesting to understand if there is a 
string field theory that generates the S-dual perturbation expansion above.
One may be able to get some hints by examining the coefficients of the
expansion of $\Pi^{-1}$ in a basis of twist fields.  Such an expression
could be obtained using the character
expansion of $P_R$
to obtain 
a formula for $\Pi^{-1}$  as an expansion  in terms of the twist operator  basis of $\cZ ( \mC_{\omega} ( G ) ) $. 
The expansion coefficients involve  the calculation of the sums
\bea 
\sum_{ R } (\dim R)^3 \chi^R ( g ) . 
\eea
These sums are some functions of $g$.  It would be interesting to  find out how these depend on the conjugacy class of $g$.

For example in $\mC ( S_3) $ it is easy to calculate  
\bea 
\Pi = 18 + 9  ( ( 1,2,3) +  ( 1,3,2 ) ),  \cr 
\Pi^{ -1} = { 1 \over 12}  - {1 \over 36 } ( ( 1,2,3) +  (1,2,3) ) . 
\eea
It would be interesting to explore this for general $\mC ( S_n) $ and other group algebras.

Dualities in the context of discrete gauge theories have been discussed in \cite{Matt2109,Matt2012}. It will be interesting to investigate potential relations between these dualities and the $S$-duality considered here.

\subsection{Finiteness relations } 

Systematic studies of the consequences of finiteness of $G$ on the 
string amplitudes of $G$-CTST, both in untwisted and twisted case, 
are interesting future directions. 
For any group $G$, with $K$ conjugacy classes, there are universal $K$-dependent finiteness relations which were described explicitly in \cite{deMelloKoch:2021lqp}. 
Requiring that these finite $K$ relations 
appear as null states of an inner product led to a discussion of the factorization puzzle in 2D/3D holography \cite{MM}. 
The inner product discussed in \cite{deMelloKoch:2021lqp} was not uniquely determined. 
It would be interesting to investigate if there is a  natural 
inner product, determined by the finiteness relations, 
possibly with additional data naturally related to $G$-CTST. As we have seen in this paper, the degeneracies of representation theoretic data (e.g. of values of dimensions of irreps) have important implications for integrality. They can be expected to play a role in $G$-dependent refinements of the finite $K$ relations.

\vskip.5cm 

\begin{center} 
	{ \bf Acknowledgements} 
\end{center} 

We are pleased to acknowledge useful conversations on the subject of this paper 
with Matt Buican, Robert de Mello Koch,  Yang Hui He, Gareth Jones,
Garreth Kemp,  Adrian Padellaro, Rajath Radhakrishnan, and Joan Simon. 
S.R.~was supported by the STFC consolidated grant ST/P000754/1 ``String Theory, Gauge Theory \& Duality” and a Visiting Professorship at the University of the Witwatersrand. E.S.~was partially supported by NSF grant PHY-2014086.

\begin{appendix} 

\section{Basics of group cohomology}\label{app:GpCoh}

Briefly, group cohomology $H^n(G,U(1))$ can be represented by
cochains $C^n(G,U(1))$, meaning maps $G^n \rightarrow U(1)$, 
which are closed in the sense
$d \omega = 0$ for 
\begin{equation}
d: \: C^n(G,U(1)) \: \longrightarrow \: C^{n+1}(G,U(1))
\end{equation}
defined by
\begin{eqnarray}
(d \omega)(g_1, \cdots, g_{n+1}) & = &
\omega(g_2, \cdots, g_{n+1}) \left[
\prod_{i=1}^n  \left( \omega(g_1, \cdots, g_{i-1}, g_i g_{i+1}, g_{i+2},
\cdots) \right)^{(-)^i} \right]
\nonumber \\
& & \hspace*{1in}  \cdot
\left( \omega(g_1, \cdots, g_n) \right)^{(-)^{n+1}},
\end{eqnarray}
modulo coboundaries, meaning the image of $d: C^{n-1}(G,U(1)) \rightarrow
C^n(G,U(1))$.

For example, $[\omega] \in H^2(G,U(1))$ are maps $\omega: G^2 \rightarrow U(1)$
such that
\begin{equation}
\frac{ \omega(g_2, g_3) }{ \omega(g_1 g_2, g_3) } \,
\frac{  \omega(g_1, g_2 g_3) }{
 \omega(g_1, g_2) } \: = \: 1,
\end{equation}
modulo equivalences
\begin{equation}
\omega(g_1, g_2) \: \sim \: \omega(g_1, g_2)  \, \frac{  b(g_1)\, b(g_2) }{
 b(g_1 g_2) }
\end{equation}
for $b: G \rightarrow U(1)$.

One can always pick cocycles so that, for example for 2-cocycles,
\begin{equation}
\omega(1,g) \: = \: 1 \: = \: \omega(g,1)
\end{equation}
for any group element $g$.  We work with such normalized cocycles in
in this paper.

\section{Characters of projective representations}\label{app:CharProj} 

In this appendix we review some basics facts and results on characters of
projective representations of finite groups that are used elsewhere
in this paper.

Perhaps the first result to recall is that, unlike characters of
ordinary group representations, characters of projective representations
are not class functions (not invariant under conjugation),
but instead obey  \cite[section 7.2, prop. 2.2]{karpilovsky}
\begin{equation} \label{eq:conj-char-again}
\chi^R(g) \: = \: \frac{ \omega(g,h^{-1}) }{ 
\omega(h^{-1}, h g h^{-1}) }
\chi^R(h g h^{-1}),
\end{equation}
as was previously mentioned in~(\ref{eq:conj-char}).

Second, these characters
vanish on non-$\omega$-regular group elements, see
e.g. \cite[section 7.2, prop. 2.2]{karpilovsky}, where an element $g \in G$
is said to be $\omega$-regular if for all $h$ commuting with $g$,
\begin{equation}
\omega(g,h) \: = \: \omega(h,g).
\end{equation}
Irreducible projective representations are in one-to-one
correspondence with $\omega$-regular conjugacy classes.

Next, we know (see e.g.~\cite[section 7.3]{karpilovsky},
\cite[section 31.1]{cr}, \cite[equ'ns (B.4), (B.20)]{Sharpe:2021srf})
\begin{eqnarray} \label{eq:char-master}
\frac{1}{|G|} \sum_{g \in G} 
\frac{ D^R(g)_{ju} D^S(g^{-1})_{ik} }{ \omega(g,g^{-1}) }
& = &
\frac{ \delta_{R,S} }{ \dim R}
\delta_{jk} \delta_{ui}, 
\end{eqnarray}
and, for $[g]$, $[h]$ both\footnote{
If either is not an $\omega$-regular conjugacy class, then the corresponding
characters vanish, and the sum equals zero.
} $\omega$-regular conjugacy classes,
\begin{eqnarray} 
\sum_R \frac{
\chi^R(g) \chi^R(h^{-1})
}{
\omega(h,h^{-1})
}
& = & \left\{ \begin{array}{cl}
0 & g, h \mbox{ not conjugate}, \\
\frac{ |G|}{|[g]|}  & g = h, \\
\frac{ \omega(a,g) }{ \omega(h,a) } \frac{ |G|}{|[g]|}
 & g = a^{-1} h a,
\end{array} \right.
\label{eq:char-master2}
\end{eqnarray}
where $R$, $S$ are irreducible projective representations
(with respect to $\omega$),
$D^R(g)$ is a matrix representing $g \in G$ in $R$, meaning
\begin{equation}
D^R(g) D^R(h) \: = \: \omega(g,h) D^R(gh),
\end{equation}
the sum in the second identity is over irreducible
projective representations,
and $|[g]|$ denotes the number of elements in a conjugacy class
containing $g$.

For use in other sections, from the expressions above one can
show (see e.g.~\cite[appendix B]{Sharpe:2021srf})
\begin{equation} \label{eq:char-om1}
\frac{1}{|G|} \sum_{g \in G} 
\frac{ \omega(a,g) \omega(g^{-1},b) }{ \omega(g,g^{-1}) } \,
\chi^R(ag) \, \chi^S(g^{-1} b)
\: = \:
\frac{ \delta_{R,S} }{\dim R} \, \omega(a,b) \, \chi^R(ab).
\end{equation}
\begin{equation} \label{eq:char-om1r}
\frac{1}{|G|} \sum_{g \in G} 
\frac{ \omega(g,a) \omega(b,g^{-1}) }{ \omega(g,g^{-1}) }
\chi^R(ga) \chi^S(bg^{-1})
\: = \: 
\frac{\delta_{R,S}}{\dim R} \omega(a,b) \chi^R(ab).
\end{equation}
\begin{equation} \label{eq:char-om2}
\frac{1}{|G|} \sum_{g \in G} \frac{ \omega(g,a) \, \omega(g^{-1},b) \,
\omega(ga, g^{-1} b) }{ \omega(g,g^{-1}) } \,
\chi^R(g a g^{-1} b) \: = \:
\frac{1}{\dim R} \, \chi^R(a) \, \chi^R(b).
\end{equation}
\begin{equation}  \label{eq:char-om2r}
\frac{1}{|G|} \sum_{g \in G} \frac{
\omega(a,g) \, \omega(b,g^{-1}) \, \omega(ag,bg^{-1}) }{ \omega(g,g^{-1}) }
\chi^R(a g b g^{-1}) \: = \: 
\frac{1}{\dim R} \, \chi^R(a) \chi^R(b).
\end{equation}
(Alternatively, by writing in terms of characters of products of $\tau$'s,
one can produce equivalent expressions without factors of $\omega$.)

Furthermore, from~(\ref{eq:char-master2}), it is straightforward to show
that
\begin{equation}  \label{eq:delta}
\delta\left( g \right)
\: = \:
\sum_R \frac{\dim R}{|G|} \chi^R\left( g \right).
\end{equation}

Let us check that this identity is well-defined under conjugation.
Using~(\ref{eq:conj-char-again}),
\begin{eqnarray}
\delta\left( h g h^{-1} \right)
& = &
\sum_R \frac{\dim R}{|G|} \chi^R\left( h g h^{-1} \right),
\\
& = &
\frac{  \omega(h^{-1}, hgh^{-1} ) }{ \omega(g,h^{-1}) }
\delta(g).
\end{eqnarray}
If $h g h^{-1} \neq 1$, then both sides vanish, so there is no ambiguity.
Similarly, if $h g h^{-1} = 1$, then $g=1$, and
\begin{equation}
\frac{  \omega(h^{-1}, hgh^{-1} ) }{ \omega(g,h^{-1}) } \: = \: 1,
\end{equation}
so again the identity is unambiguous.

In passing, for the projector $P_R$ given in equation~(\ref{eq:proj-defn}),
note that this implies
\begin{eqnarray}
\delta\left( P_R \right) & = &
\frac{\dim R}{|G|} \sum_{g \in G} 
\frac{ \chi^R(g^{-1}) }{ \omega(g,g^{-1}) } \delta(g),
\\
& = &
\frac{ \dim R}{|G|^2} \sum_{g \in G} 
\frac{ \chi^R(g^{-1}) }{ \omega(g,g^{-1}) }
\sum_S (\dim S) \chi^S(g),
\end{eqnarray}
but from~(\ref{eq:char-om1}), one has
\begin{equation}
\sum_{g \in G} \frac{ \chi^R(g) \chi^S(g^{-1}) }{ \omega(g,g^{-1}) }
\: = \: |G| \delta_{RS},
\end{equation}
hence
\begin{eqnarray}
\delta\left( P_R \right) & = &
\sum_S \frac{ (\dim R) (\dim S) }{|G|^2} |G| \delta_{RS},
\\
& = &
\frac{ (\dim R)^2 }{|G|}.    \label{eq:delta-proj}
\end{eqnarray}

Another identity that will be useful involves the
handle creation operator $\Pi$ given in~(\ref{eq:handle-op-id}).
Using~(\ref{eq:char-om1}), first note that
\begin{equation}
\chi^S(P_R) \: = \:
\frac{\dim R}{|G|} \sum_{g \in G} \frac{ \chi^R(g^{-1}) \chi^S(g) }{ 
\omega(g,g^{-1}) }
\: = \:
(\dim R) \delta_{R,S}.
\end{equation}
Then,
\begin{eqnarray}
\chi^S(\Pi) & = &
\sum_R \left( \frac{ |G| }{ \dim R} \right)^2 \chi^S(P_R),
\\
& = &
\frac{ |G|^2 }{ \dim R}.
\label{eq:char-handle-op}
\end{eqnarray}

One of the consequences of the fact that characters of projective
representations are not invariant under conjugation is that,
unlike characters of ordinary representations for which
\begin{equation}
\chi^R(gh) \: = \chi^R(hg)
\end{equation}
characters of projective representations instead have the property
\begin{equation}
\chi^R(gh) \: \neq \: \chi^R(hg)
\end{equation}
in general.
For example, for representations of $D_4$ with nontrivial discrete torsion,
then from \cite[section 4.5]{Sharpe:2021srf} and references therein,
\begin{equation}
\chi_r(b = baa) \: \neq \: \chi_r(aba = bz).
\end{equation}
For example, in the notation of that reference,
\begin{equation}
\chi_1(b) \: = \: 1+i,
\: \: \:
\chi_1(bz) \: = \: 1-i.
\end{equation}
In fact, we can derive a general relation between
$\chi^R(gh)$ and $\chi^R(hg)$ as follows.
In principle,
\begin{equation}
\chi^R(g) \: = \: {\rm Tr} \, \rho_R(g),
\end{equation}
where $\rho_R(g)$ is a matrix representing $g$.  Now,
\begin{eqnarray}
\chi^R(gh) & = & {\rm Tr}\, \rho_R(gh),
\\
& = &
\omega(g,h)^{-1}
{\rm Tr} \, \rho_R(g) \rho_R(h),
\\
& = &
\omega(g,h)^{-1}
{\rm Tr} \, \rho_R(h) \rho_R(g),
\\
& = &
\frac{\omega(h,g) }{ \omega(g,h) }
{\rm Tr}\, \rho_R(hg),
\\
& = & 
\frac{\omega(h,g) }{ \omega(g,h) }
\chi^R(hg).
\end{eqnarray}

As a consistency check, we claim that $\delta(gh) = \delta(hg)$.
Now, from~(\ref{eq:delta}),
\begin{eqnarray}
\delta(gh) & = & \sum_R \frac{\dim R}{|G|} \chi^R(gh),
\\
& = & \frac{\omega(h,g)}{\omega(g,h)} 
\sum_R \frac{\dim R}{|G|} \chi^R(hg),
\\
& = &
 \frac{\omega(h,g)}{\omega(g,h)} 
\delta(hg).
\end{eqnarray}
Now, if $gh \neq 1$, then $hg \neq 1$, so both sides of the relation above
vanish, and in particular, $\delta(gh) = \delta(hg) = 0$.
Suppose instead that $gh = 1$, so that $\delta(gh) = 1$.
In this case, $h = g^{-1}$, and from
\begin{equation}
(d \omega)(g,g^{-1},g) \: = \: 1,
\end{equation}
we have
\begin{equation}
\omega(g,g^{-1}) = \omega(g^{-1},g).
\end{equation}
Thus, if $gh = 1$, then $\delta(gh) = \delta(hg) = 1$,
so for all $g$ and $h$, $\delta(gh) = \delta(hg)$.

\section{Two-dimensional Dijkgraaf-Witten theory}
\label{app:2ddw-results}

In this appendix we collect some technical results on 
two-dimensional twisted Dijkgraaf-Witten theory that are used in the
main text.  Although we have not located a complete set of prior references,
we believe these results were known previously; we include them and their
derivations here for completeness and to make the detailed arguments of
the main text convincing.

\subsection{Partition functions}\label{PFsDets}

In this section we will compute genus $g$ partition functions of
two-dimensional Dijkgraaf-Witten theory with discrete torsion,
in the same style as the analysis of \cite[section 2]{deMelloKoch:2021lqp}
to include discrete torsion.
Now, to be clear, these partition functions have been computed
previously in the literature, see for example
\cite{Gardiner:2020vjp} in the physics literature for a recent
computation in two-dimensional
Dijkgraaf-Witten theory specifically,
\cite[appendix C.1]{Komargodski:2020mxz} for a recent review of results on
partition functions of 2d TQFTs,
and in the math literature, see for example
\cite{frobenius1,frschur,mednyh,mulaseyu,Snyder:2007ns} for partition functions
and one-point functions in cases without\footnote{
Partition functions including discrete torsion have certainly
been computed previously in the physics literature, see
e.g. \cite{Dijkgraaf:1989pz}.  We include such computations here
for completeness.
Our expectation is that partition functions including discrete torsion
were also computed, albeit in different language, in the mathematics
literature in the same era as
\cite{frobenius1,frschur}, though we have not been able to find a specific
mathematics reference.
} discrete torsion,
where
these are given as the orbifold Euler characteristics
of the moduli space of flat $G$ bundles,
\begin{eqnarray}
\chi_{\rm orb}\left( {\cal M}_G(\Sigma) \right) & = &
\chi_{\rm orb}\left( {\rm Hom}(\pi_1(\Sigma), G) / G \right),
\\
& = & 
\frac{ | {\rm Hom}(\pi_1(\Sigma), G) |}{|G|},
\\
& = &
\sum_{\rho \in {\rm Hom}(\pi_1(\Sigma),G)/G} 
\frac{1}{| {\rm Aut}(\rho) |},
\\
& = &
\sum_R \left( \frac{ \dim R}{|G|} \right)^{\chi(\Sigma)}.
\end{eqnarray}
(We do not claim to give a complete list of references,
but merely list a few representative examples; additional references are
given in e.g.~\cite{mo-z,mulaseyu,Snyder:2007ns}.)
Also, in passing, an alternative computation of the same result is
given in section~\ref{sect:handle-op}.

First, we consider the genus-one partition function.
Using~(\ref{eq:char-master}) it is straightforward to check
\begin{equation} 
\sum_{g_1, g_2 \in G}
\frac{ D^R(g_1)_{ab} D^R(g_2)_{bc} D^R(g_1^{-1})_{cd} D^R(g_2^{-1})_{de}
}{
\omega(g_1, g_1^{-1}) \omega(g_2, g_2^{-1})
}
\: = \:
\left( \frac{ |G| }{ \dim R } \right)^2 \delta_{ae}.
\end{equation}
(Compare \cite[equ'n (2.4)]{deMelloKoch:2021lqp}.)
From this one immediately derives
\begin{equation}  \label{eq:prod4}
\sum_{g_1, g_2} D^R_{ae}\left( [g_1, g_2] \right)
\frac{ 
\omega(g_1, g_2) \omega(g_1^{-1}, g_2^{-1})
\omega(g_1 g_2, g_1^{-1} g_2^{-1})
}{
\omega( g_1, g_1^{-1}) \omega(g_2, g_2^{-1})
}
\: = \:
\left( \frac{ |G| }{ \dim R } \right)^2 \delta_{ae},
\end{equation}
where
\begin{equation}
[g_1, g_2] \: = \: g_1 g_2 g_1^{-1} g_2^{-1}.
\end{equation}
In particular,
\begin{equation} \label{eq:genus1:charsum}
\sum_{g_1, g_2} \chi^R\left( [g_1, g_2] \right)
\frac{ 
\omega(g_1, g_2) \omega(g_1^{-1}, g_2^{-1})
\omega(g_1 g_2, g_1^{-1} g_2^{-1})
}{
\omega( g_1, g_1^{-1}) \omega(g_2, g_2^{-1})
}
\: = \:
\left( \frac{ |G| }{ \dim R } \right)^2
\dim R.
\end{equation}

By multiplying in
\begin{equation}
\frac{
(d \omega)(g_2, g_1, g_1^{-1}) 
}{
(d\omega)(g_1 g_2, g_1^{-1}, g_2^{-1})
} \: = \: 1,
\end{equation}
one finds that in the special case $g_1 g_2 = g_2 g_1$,
\begin{equation}
\frac{ 
\omega(g_1, g_2) \omega(g_1^{-1}, g_2^{-1})
\omega(g_1 g_2, g_1^{-1} g_2^{-1})
}{
\omega( g_1, g_1^{-1}) \omega(g_2, g_2^{-1})
}
\: = \:
\frac{
\omega(g_1, g_2)
}{ 
\omega(g_2, g_1)
}.
\end{equation}

Using (\ref{eq:delta}),
\begin{equation}
\delta\left( [g_1, g_2] \right)
\: = \:
\sum_R \frac{\dim R}{|G|} \chi^R\left( [g_1, g_2] \right).
\end{equation}

Assembling these pieces, we have that the genus-one partition function
(with discrete torsion) is given by
\begin{eqnarray}
Z_{g=1} & = &
\frac{1}{|G|} \sum_{g_1, g_2} \delta\left( [g_1, g_2] \right)
\frac{ \omega(g_1, g_2) }{ \omega(g_2, g_1) },
\\
& = &
\frac{1}{|G|} \sum_{g_1, g_2} \delta\left( [g_1, g_2] \right)
\frac{ 
\omega(g_1, g_2) \omega(g_1^{-1}, g_2^{-1})
\omega(g_1 g_2, g_1^{-1} g_2^{-1})
}{
\omega( g_1, g_1^{-1}) \omega(g_2, g_2^{-1})
},
\\
& = &
\frac{1}{|G|} \sum_{g_1, g_2} \left[
\sum_R \frac{\dim R}{|G|} \chi^R\left( [g_1, g_2] \right)
\right]
\frac{     
\omega(g_1, g_2) \omega(g_1^{-1}, g_2^{-1})
\omega(g_1 g_2, g_1^{-1} g_2^{-1})
}{
\omega( g_1, g_1^{-1}) \omega(g_2, g_2^{-1})
},
\nonumber \\
& = &
\sum_R \frac{\dim R}{|G|^2} \left[ 
\sum_{g_1, g_2} \chi^R\left( [g_1, g_2] \right)
\frac{     
\omega(g_1, g_2) \omega(g_1^{-1}, g_2^{-1})
\omega(g_1 g_2, g_1^{-1} g_2^{-1})
}{
\omega( g_1, g_1^{-1}) \omega(g_2, g_2^{-1})
}
\right],
\\
& = & \sum_R \frac{ \dim R}{|G|^2}
\left( \frac{ |G| }{\dim R} \right)^2 \dim R,
\\
& = &
\sum_R (1),
\end{eqnarray}
where we used~(\ref{eq:genus1:charsum}).
This recovers the result in \cite[equ'n (6.40)]{Dijkgraaf:1989pz}.

Next, we compute the partition functions on Riemann surfaces of
general genus.
We will follow the notation of \cite{Aspinwall:2000xv}.
Consider a Riemann surface of genus $g$,
with insertions defined by group elements
$a_i$, $b_i$, $i \in \{1, \cdots, g\}$.
Define $\gamma_i = [ a_i, b_i]$,
and
\begin{equation}
        X \: = \: \left[ \prod_i \omega(a_i, a_i^{-1})
        \prod_i \omega(b_i,b_i^{-1}) \right]^{-1}.
\end{equation}

Then, from~(\ref{eq:char-master}), we have that
\begin{eqnarray}
\lefteqn{
\sum_{a_i, b_i} \frac{
D^R(a_1) D^R(b_1) D^R(a_1^{-1}) D^R(b_1^{-1}) \cdots D^R(b_g^{-1})
}{
\omega(a_1, a_1^{-1}) \omega(b_1, b_1^{-1}) \cdots
\omega(b_g, b_g^{-1})
}
\: = \:
\left( \frac{ |G| }{ \dim R } \right)^{2g} I,
} \\
& = & \sum_{a_i, b_i} D^R\left( \gamma_1 \cdots \gamma_g \right)
X
        \omega(a_1, b_1) \omega(a_1 b_1, a_1^{-1}) 
        \omega(a_1 b_1 a_1^{-1}, b_1^{-1})
        \omega(\gamma_1, a_2) \omega(\gamma_1 a_2, b_2)
        \omega(\gamma_1 a_2 b_2, a_2^{-1})
\nonumber \\
& & \hspace*{1in} \cdot
        \omega(\gamma_1 a_2 b_2 a_2^{-1}, b_2^{-1})
        \omega(\gamma_1 \gamma_2, a_3) \cdots
        \omega(\gamma_1 \cdots \gamma_{g-1} a_g b_g a_g^{-1}, b_g^{-1}).
\end{eqnarray}

Now, the phase factor assigned by discrete torsion to a genus $g$
Riemann surface is \cite[equ'n (15)]{Aspinwall:2000xv} (see also
\cite{Bantay:2000eq})
\begin{eqnarray}
\lefteqn{
        \epsilon_g(a_i, b_i) \: \equiv \:
        X
        \omega(a_1, b_1) \omega(a_1 b_1, a_1^{-1})
        \omega(a_1 b_1 a_1^{-1}, b_1^{-1})
        \omega(\gamma_1, a_2) \omega(\gamma_1 a_2, b_2)
        \omega(\gamma_1 a_2 b_2, a_2^{-1})
} \nonumber \\
& & \hspace*{1in} \cdot
        \omega(\gamma_1 a_2 b_2 a_2^{-1}, b_2^{-1})
        \omega(\gamma_1 \gamma_2, a_3) \cdots
        \omega(\gamma_1 \cdots \gamma_{g-1} a_g b_g a_g^{-1}, b_g^{-1}).
\end{eqnarray}
Thus, we can write the expression above as
\begin{equation}
        \sum_{a_i, b_i} D^R\left( \gamma_1 \cdots \gamma_g \right)
        \epsilon_g(a_i, b_i) 
        \: = \:
        \left( \frac{ |G| }{ \dim R } \right)^{2g} I.
\end{equation}

In particular,
\begin{eqnarray}
\sum_{a_i, b_i} \chi^R\left( \gamma_1 \cdots \gamma_g \right)
\epsilon_g(a_i, b_i)
& = &
\left( \frac{ |G| }{ \dim R } \right)^{2g} (\dim R).
\label{eq:genusg:charsum}
\end{eqnarray}

Applying the identity~(\ref{eq:delta})
\begin{equation}
\delta(\gamma_1 \cdots \gamma_g) \: = \:
\sum_R \frac{\dim R}{|G|} \chi^R(\gamma_1 \cdots \gamma_g),
\end{equation}
we then compute
\begin{eqnarray}
Z_g & = &
\frac{1}{|G|} \sum_{a_i, b_i} \delta\left( \prod_i \gamma_i\right) 
\epsilon_g(a_i, b_i),
\\
& = &
\frac{1}{|G|} \sum_{a_i, b_i} \left[
\sum_R \frac{\dim R}{|G|} \chi^R(\gamma_1 \cdots \gamma_g)
\right]
\epsilon_g(a_i, b_i), 
\\
& = &
\frac{1}{|G|} \sum_R \frac{\dim R}{|G|} \Biggl[
 \sum_{a_i, b_i} \chi^R(\gamma_1 \cdots \gamma_g)
\epsilon_g(a_i, b_i)
\Biggr],
\\
& = &
\frac{1}{|G|} \sum_R \frac{\dim R}{|G|} \left[
\left( \frac{ |G| }{ \dim R } \right)^{2g} (\dim R)
\right],
\\
& = &
\sum_R \left( \frac{ |G| }{ \dim R } \right)^{2g-2},
\label{eq:partfn}
\end{eqnarray}
where we have used equation~(\ref{eq:genusg:charsum}).

\subsection{Handle creation operator}
\label{sect:handlecreation}

In this section, we will describe the handle creation operator
in the presence of discrete torsion, and its basic properties.

Without discrete torsion, the handle creation operator is
\cite[equ'n (6.23)]{deMelloKoch:2021lqp}
\begin{equation}
\Pi \: = \: \sum_{g_1, g_2 \in G} \tau_{g_1} \tau_{g_2} \tau_{g_1}^{-1} 
\tau_{g_2}^{-1},
\end{equation}
and it is claimed that
\begin{equation}   \label{eq:handle-op-id-without-dt}
\Pi \: = \: \sum_R \left( \frac{|G|}{\dim R} \right)^2 P_R,
\end{equation}
for $P_R$ the projection operator.

As a consistency test, note this implies
\begin{equation}  \label{eq:predict-wodt}
\sum_{g_1, g_2 \in G} D^S\left(g_1 g_2 g_1^{-1} g_2^{-1} \right)
\: = \:
\sum_R \left( \frac{|G|}{\dim R} \right)^2 D^S(P_R).
\end{equation}

Let us check that this implication is correct for every
irreducible representation $S$.
First, from~(\ref{eq:prod4}), in the absence of discrete torsion,
we have
\begin{equation}
\sum_{g_1, g_2 \in G} D^S\left(g_1 g_2 g_1^{-1} g_2^{-1} \right)
\: = \:
\left( \frac{|G|}{\dim S} \right)^2 I.
\end{equation}
Now,
\begin{equation}
P_R \: = \: \frac{\dim R}{|G|} \sum_{g \in G} \chi^R(g^{-1}) \tau_{g},
\end{equation}
hence
\begin{eqnarray}
D^S(P_R) & = &
\frac{\dim R}{|G|} \sum_{g \in G} \chi_R(g^{-1}) T^S(g),
\\
& = &  \delta_{R,S}  I
\mbox{  using }(\ref{eq:char-master}),
\end{eqnarray}
hence
\begin{eqnarray}
\sum_{g_1, g_2 \in G} D^S\left(g_1 g_2 g_1^{-1} g_2^{-1} \right)
& = &
\sum_R \left( \frac{|G|}{\dim R} \right)^2  \delta_{R,S}  I,
\\
& = &
\sum_R \left( \frac{|G|}{\dim R} \right)^2 D^S(P_R),
\end{eqnarray}
confirming~(\ref{eq:predict-wodt}.
Since this holds for any irreducible representation $S$, we take this
as a confirmation of the handle creation operator
identity~(\ref{eq:handle-op-id-without-dt}).

Now, let us turn to the case with discrete torsion.

Here, we define the handle creation operator to be
\begin{eqnarray}\label{HandComb} 
\Pi & = &
\sum_{ g_1 , g_2 \in G } \tau_{ g_1 } \tau_{ g_2} \tau_{ g_1}^{-1} \tau_{ g_2}^{-1} \cr 
& = &
\sum_{g_1, g_2 \in G} 
\frac{ 
\omega(g_1, g_2) \omega(g_1^{-1}, g_2^{-1})
\omega(g_1 g_2, g_1^{-1} g_2^{-1})
}{
\omega( g_1, g_1^{-1}) \omega(g_2, g_2^{-1})
}
\tau_{g_1 g_2 g_1^{-1} g_2^{-1}},
\end{eqnarray}
and we claim that
\begin{equation} \label{eq:handle-op-id}
\Pi \: = \: \sum_R \left( \frac{|G|}{\dim R} \right)^2 P_R.
\end{equation}

As a consistency check, this implies that
\begin{equation}  \label{eq:predict-wdt}
\sum_{g_1, g_2 \in G} 
\frac{ 
\omega(g_1, g_2) \omega(g_1^{-1}, g_2^{-1})
\omega(g_1 g_2, g_1^{-1} g_2^{-1})
}{
\omega( g_1, g_1^{-1}) \omega(g_2, g_2^{-1})
}
D^S\left( g_1 g_2 g_1^{-1} g_2^{-1} \right)
\: = \:
\sum_R \left( \frac{|G|}{\dim R} \right)^2 D^S(P_R).
\end{equation}

Let us check that this implication is correct for every irreducible
representation $S$.
First, from~(\ref{eq:prod4}),
we have
\begin{equation}
\sum_{g_1, g_2 \in G} 
\frac{ 
\omega(g_1, g_2) \omega(g_1^{-1}, g_2^{-1})
\omega(g_1 g_2, g_1^{-1} g_2^{-1})
}{
\omega( g_1, g_1^{-1}) \omega(g_2, g_2^{-1})
}
D^S\left( g_1 g_2 g_1^{-1} g_2^{-1} \right)
\: = \:
\left( \frac{|G|}{\dim R} \right)^2 I.
\end{equation}
Now, with discrete torsion,
\begin{eqnarray}
P_R & = & \frac{\dim R}{|G|} \sum_{g \in G} 
\frac{ \chi^R(g^{-1}) }{ \omega(g,g^{-1}) } g,
\end{eqnarray}
so
\begin{eqnarray}
D^S(P_R) & = & \frac{\dim R}{|G|} \sum_{g \in G} 
\frac{ \chi^R(g^{-1}) }{ \omega(g,g^{-1}) } D^S(g),
\\
& = &
\delta_{R,S} I
\mbox{  using }(\ref{eq:char-master}),
\end{eqnarray}
hence
\begin{eqnarray}
\lefteqn{
\sum_{g_1, g_2 \in G} 
\frac{ 
\omega(g_1, g_2) \omega(g_1^{-1}, g_2^{-1})
\omega(g_1 g_2, g_1^{-1} g_2^{-1})
}{
\omega( g_1, g_1^{-1}) \omega(g_2, g_2^{-1})
}
D^S\left( g_1 g_2 g_1^{-1} g_2^{-1} \right)
} \nonumber \\
& \hspace*{1.5in}  = &
\sum_R \left( \frac{|G|}{\dim R} \right)^2 
\delta_{R,S} I,
\\
& \hspace*{1.5in}  = &
\sum_R \left( \frac{|G|}{\dim R} \right)^2 
D^S(P_R)
\end{eqnarray}
confirming~(\ref{eq:predict-wdt}).
Since this holds for any irreducible representation $S$, we take this
as a confirmation of the handle creation operator
identity~(\ref{eq:handle-op-id}).

\subsection{Handle creation operator identities}

In this section, we will describe some handle-operator creation identities.

First, we claim that
\begin{equation}
\delta\left( \Pi^n T_{[g]} \right) \: = \: \sum_R \left(
\frac{ |G| }{ \dim R} \right)^{2n-1}
\chi^R(g).
\end{equation}

To this end, recall the identity~(\ref{eq:handle-op-id})
\begin{equation}
\Pi \: = \: \sum_R \left( \frac{|G|}{\dim R} \right)^2 P_R,
\end{equation}
where $P_R$ is the projector given by
\cite[equ'n (2.43)]{Sharpe:2021srf}
\begin{equation}
P_R \: = \: \frac{\dim R}{|G|} \sum_{k \in G} \frac{ \chi^R(k^{-1}) }{
\omega(k,k^{-1}) } \tau_k,
\end{equation}
hence
\begin{equation}
\Pi^n \: = \: \sum_R \left(  \frac{|G|}{\dim R} \right)^{2n} P_R,
\end{equation}
and \cite[equ'n (2.17)]{Sharpe:2021srf}
\begin{equation}
T_{[g]} \: = \: \frac{1}{|G|} 
\sum_{h \in G} \frac{
\omega(h,g) \omega(hg,h^{-1}) }{ \omega(h,h^{-1}) }
\tau_{ h g h^{-1} }.
\end{equation}
Thus,
\begin{eqnarray}
\delta\left( \Pi^n T_{[g]} \right) 
& = &
\sum_R \left(  \frac{|G|}{\dim R} \right)^{2n}
 \frac{\dim R}{|G|} \sum_{k \in G} \frac{ \chi^R(k^{-1}) }{
\omega(k,k^{-1}) }
 \frac{1}{|G|} 
\sum_{h \in G} \frac{
\omega(h,g) \omega(hg,h^{-1}) }{ \omega(h,h^{-1}) }
\nonumber \\
& & \hspace*{2in} \cdot
\omega(k, h g h^{-1}) \delta(k h g h^{-1}),
\end{eqnarray}
where we have used the fact that
\begin{equation}
\tau_g \tau_h \: = \: \omega(g,h) \tau_{gh}.
\end{equation}
Using
\begin{equation}
\frac{ (d \omega)(kh, g, h^{-1}) }{
(d \omega)(k, hg, h^{-1}) \, (d\omega)(k, h, g) }
\: = \: 1
\end{equation}
we have
\begin{equation}  \label{eq:gpidentity1}
\frac{ \omega(h,g) \, \omega(hg, h^{-1}) \, \omega(k, h g h^{-1}) }{
\omega(h,h^{-1}) }
\: = \:
\frac{ \omega(k,h) \, \omega(g,h^{-1}) \, \omega(kh, gh^{-1}) }{
\omega(h,h^{-1}) },
\end{equation}
hence
\begin{eqnarray}
\delta\left( \Pi^n T_{[g]} \right) 
& = &
\sum_R   \left( \frac{|G|}{\dim R} \right)^{2n-1}
 \sum_{k \in G} \frac{ \chi^R(k^{-1}) }{
\omega(k,k^{-1}) }
 \frac{1}{|G|} 
\sum_{h \in G}
\frac{ \omega(k,h) \, \omega(g,h^{-1}) \, \omega(kh, gh^{-1}) }{
\omega(h,h^{-1}) }
\nonumber \\
& & \hspace*{2in} \cdot
\sum_S \frac{\dim S}{|G|} \chi^R(k h g h^{-1}).
\end{eqnarray}
Using~(\ref{eq:char-om2r}),
\begin{eqnarray}
\delta\left( \Pi^n T_{[g]} \right) 
& = &
\sum_R \left(  \frac{|G|}{\dim R} \right)^{2n-1}
 \sum_{k \in G} \frac{ \chi^R(k^{-1}) }{
\omega(k,k^{-1}) }
\sum_S \frac{\dim S}{|G|}
\frac{1}{\dim S} \chi^S(k) \chi^S(g),
\\
& = &
\sum_R \left(  \frac{|G|}{\dim R} \right)^{2n-1}
\sum_S \delta_{R,S} \chi^S(g),
\\
& = & |G| \sum_R \left(  \frac{|G|}{\dim R} \right)^{2n-2}
\frac{ \chi^R(g) }{ \dim R},
\end{eqnarray}
using~(\ref{eq:char-om1}).

Next, we compute
\begin{eqnarray}
\delta\left( \Pi^n T_{[g_1]} T_{[g_2]} \right)
& = &
\sum_R \left( \frac{|G|}{\dim R} \right)^{2n} \frac{\dim R}{|G|}
\sum_k \frac{ \chi^R(k^{-1}) }{ \omega(k,k^{-1})}
\frac{1}{|G|} \sum_{h_1} \frac{
\omega(h_1, g_1) \omega(h_1 g_1, h_1^{-1}) }{ \omega(h_1, h_1^{-1}) }
\nonumber \\
& & \hspace*{1.5in} \cdot
\frac{1}{|G|} \sum_{h_2}  \frac{
\omega(h_2, g_2) \omega(h_2 g_2, h_2^{-1}) }{ \omega(h_2, h_2^{-1}) }
\nonumber \\
& & \hspace*{1.5in} \cdot
\omega(k, h_1 g_1 h_1^{-1}) \omega(k h_1 g_1 h_1^{-1}, h_2 g_2 h_2^{-1})
\nonumber \\
& & \hspace*{1.5in} \cdot
\sum_S \frac{\dim S}{|G|} \chi^S\left( k h_1 g_1 h_1^{-1} h_2 g_2 h_2^{-1} 
\right).
\end{eqnarray}
Using~(\ref{eq:gpidentity1}) and (\ref{eq:char-om2r}), this reduces to
\begin{eqnarray}
\delta\left( \Pi^n T_{[g_1]} T_{[g_2]} \right)
& = &
\sum_R \left(  \frac{|G|}{\dim R}  \right)^{2n-1}
\sum_k \frac{ \chi^R(k^{-1}) }{ \omega(k,k^{-1})}
\frac{1}{|G|} \sum_{h_1} \frac{
\omega(h_1, g_1) \omega(h_1 g_1, h_1^{-1}) }{ \omega(h_1, h_1^{-1}) }
\nonumber \\
& & \hspace*{1.5in} \cdot
\omega(k, h_1 g_1 h_1^{-1})
\sum_S \frac{\dim S}{|G|} 
\frac{ \chi^R\left( k h_1 g_1 h_1^{-1} \right) \chi^S(g_2) }{ \dim S }.
\nonumber
\end{eqnarray}
Modulo the factor of $\chi^S(g_2)$, this now essentially reduces to the
previous computation.  Using~(\ref{eq:gpidentity1}) and (\ref{eq:char-om2r})
again, we have
\begin{equation}
\delta\left( \Pi^n T_{[g_1]} T_{[g_2]} \right)
\: = \:
\sum_R \left(   \frac{|G|}{\dim R} \right)^{2n-1}
\sum_k \frac{ \chi^R(k^{-1}) }{ \omega(k,k^{-1})}
\sum_S \frac{\dim S}{|G|} 
\frac{ \chi^S(k) \chi^S(g_1) \chi^S(g_2) }{ (\dim S)^2 },
\nonumber
\end{equation}
and using~(\ref{eq:char-om1}), we have
\begin{equation}
\delta\left( \Pi^n T_{[g_1]} T_{[g_2]} \right)
\: = \:
|G| \sum_R \left( \frac{ |G| }{ \dim R } \right)^{2n-2}
\left( \frac{ \chi^R(g_1) }{ \dim R } \right)
\left( \frac{ \chi^R(g_2) }{ \dim R} \right).
\end{equation}

At this point, it is straightforward to derive an analogous expression
for cases with more factors of $T_{[g]}$.  Define
\begin{equation}
\gamma_i \: = \: h_i g_i h_i^{-1},
\: \: \:
A_i \: = \: \frac{1}{|G|} \sum_{h_i}
\frac{\omega(h_i, g_i) \omega(h_i g_i, h_i^{-1}) }{ \omega(h_i, h_i^{-1}) },
\end{equation}
we have
\begin{eqnarray}
\delta\left( \Pi^n T_{[g_1]} \cdots  T_{[g_m]} \right)
& = &
\sum_R \left( \frac{|G|}{\dim R} \right)^{2n-1}
 \sum_k \frac{ \chi^R(k^{-1}) }{ \omega(k,k^{-1})}
A_1 \cdots A_m 
\nonumber \\
& & \hspace*{0.5in} \cdot
\omega(k, \gamma_1) \omega(k \gamma_1, \gamma_2) \cdots
\omega(k \gamma_1 \gamma_2 \cdots \gamma_{m-1}, \gamma_m)
\nonumber \\
& & \hspace*{0.5in} \cdot
\sum_S \frac{\dim S}{|G|} \chi^S\left( k \gamma_1 \gamma_2 \cdots \gamma_m
\right).
\end{eqnarray}
Applying~(\ref{eq:gpidentity1}) and (\ref{eq:char-om2r}) to
perform the sum over $h_m$, this becomes
\begin{eqnarray}
\delta\left( \Pi^n T_{[g_1]} \cdots  T_{[g_m]} \right)
& = &
\sum_R \left( \frac{|G|}{\dim R} \right)^{2n-1}
 \sum_k \frac{ \chi^R(k^{-1}) }{ \omega(k,k^{-1})}
A_1 \cdots A_{m-1}
\nonumber \\
& & \hspace*{0.5in} \cdot
\omega(k, \gamma_1) \omega(k \gamma_1, \gamma_2) \cdots
\omega(k \gamma_1 \gamma_2 \cdots \gamma_{m-2}, \gamma_{m-1})
\nonumber \\
& & \hspace*{0.5in} \cdot
\sum_S \frac{\dim S}{|G|}
\chi^S\left( k \gamma_1 \gamma_2 \cdots \gamma_{m-1} \right)
\frac{\chi^S(g_m)}{\dim S}.
\end{eqnarray}
Iterating that procedure, and then using (\ref{eq:char-om1}), we find
\begin{eqnarray}
\lefteqn{
\delta\left( \Pi^n T_{[g_1]} \cdots  T_{[g_m]} \right)
} \nonumber \\
& = &
\sum_R \left( \frac{|G|}{\dim R} \right)^{2n-1}
 \sum_k \frac{ \chi^R(k^{-1}) }{ \omega(k,k^{-1})}
\nonumber \\  
& & \hspace*{0.5in} \cdot
\sum_S \frac{\dim S}{|G|}
\chi^S(k) \frac{\chi^S(g_1)}{\dim S} 
\frac{ \chi^S(g_2) }{\dim S} \cdots
\frac{ \chi^S(g_m) }{\dim S},
\\
& = &
|G| \sum_R \left( \frac{ |G| }{\dim R} \right)^{2n-2}
\left( \frac{\chi^R(g_1)}{\dim R} \right)
\left( \frac{\chi^R(g_2)}{\dim R} \right)
\cdots
\left( \frac{\chi^R(g_m)}{\dim R} \right).
\label{eq:final-handle-id}
\end{eqnarray}

Since this is linear in each factor, this immediately implies that
for $S_1, \cdots, S_m$ any elements of the center of the group algebra,
\begin{eqnarray}
\lefteqn{
\delta\left( \Pi^n S_1 \cdots  S_m \right)
} \nonumber \\
& = &
|G| \sum_R \left( \frac{ |G| }{\dim R} \right)^{2n-2}
\left( \frac{\chi^R(S_1)}{\dim R} \right)
\left( \frac{\chi^R(S_2)}{\dim R} \right)
\cdots
\left( \frac{\chi^R(S_m)}{\dim R} \right).
\label{eq:final-handle-id2}
\end{eqnarray}

As a consistency check, note that if $S_1 = \Pi$, say,
then from~(\ref{eq:char-handle-op}), we have
\begin{eqnarray}
\lefteqn{
\delta\left( \Pi^n S_1 \cdots S_m \right)
} \nonumber \\
& = &
|G| \sum_R \left( \frac{ |G| }{\dim R} \right)^{2n-2}
\left(  \frac{ |G| }{\dim R} \right)^2
\left( \frac{\chi^R(S_2)}{\dim R} \right)
\cdots
\left( \frac{\chi^R(S_m)}{\dim R} \right),
\\
& = &
|G| \sum_R \left( \frac{ |G| }{\dim R} \right)^{2n}
\left( \frac{\chi^R(S_2)}{\dim R} \right)
\cdots
\left( \frac{\chi^R(S_m)}{\dim R} \right),
\\
& = &
\delta\left( \Pi^{n+1} S_2 \cdots S_m \right).
\end{eqnarray}

\end{appendix}

\end{document}